\documentclass[a4paper,12pt]{scrartcl}
\usepackage[utf8]{inputenc}	
\usepackage[english]{babel}
\usepackage[T1]{fontenc}
\usepackage{amsmath}
\usepackage{amsfonts}
\usepackage{graphicx}
\usepackage{lmodern}
\usepackage{scrlayer-scrpage} 
\usepackage{hyperref}
\usepackage{xurl}
\usepackage[left=25mm, right=25mm, top=25mm, bottom=20mm]{geometry}
\usepackage{caption}
\usepackage{subcaption}
\usepackage{datetime}
\usepackage{xcolor}
\usepackage{subcaption} 
\usepackage{siunitx}
\usepackage{upgreek}
\usepackage{setspace}
\setkomafont{section}{\large}
\setkomafont{subsection}{\large}
\setkomafont{subsubsection}{\normalsize}
\usepackage{newtxtext}
\usepackage{newtxmath}
\setkomafont{disposition}{\normalfont\bfseries}
\setcapindent{0pt}
\newdateformat{usformat}{\monthname[\THEMONTH] \THEDAY, \THEYEAR}

\begin{document}
\author{Hauke Heyen$^1$, Michael Vogel$^{2,3}$, Florian Gossing$^{2,3}$,\\ Jakob Walowski$^1$, Karin Dahmen$^4$, Jeffrey McCord$^{2,3}$,\\ Markus Münzenberg$^1$}
\title{Disentangling topological and anomalous Hall contributions of skyrmions using Kerr microscopy and thermal transport measurements}
\date{\usformat\today}
\maketitle
\begin{center}
$^{1}$Institute of Physics, University of Greifswald, Felix-Hausdorff-Straße 6, 17489 Greifswald, Germany
\end{center}
\begin{center}
$^2$Department of Materials Science, Kiel University, Kaiserstraße 2, 24143 Kiel, Germany
\end{center}
\begin{center}
$^3$Kiel Nano, Surface and Interface Science (KiNSIS), Kiel University, Kiel, Germany
\end{center}
\begin{center}
$^4$Department of Physics, The Grainger College of Engineering, 1110 West Green Street, Urbana, IL 61801-3003, USA
\end{center}

\section{Abstract}
The topological Hall effect is a valuable tool to indicate the presence of topologically protected magnetic structures. In this work, we present topological Hall effect measurements originating from topologically protected skyrmions in Ta/CoFeB/MgO single-layer thick films with a one nanometer thick magnetic layer. The simultaneous occurrence of the small topological Hall effect and the dominating anomalous Hall effect in this material system makes direct detection challenging as compared to bulk or multilayer systems. In electronic transport measurements, both effects' contributions impact electron trajectories in the same way, overlapping in the measurement signal, and require disentanglement. Magneto-optical Kerr microscopy was used to image the surface magnetization, enabling the separation of the topological Hall effect from other Hall effect contributions. These measurements reveal a topological Hall resistivity of $249(18)\,$\si{\pico\ohm\meter} for Ta/CoFeB/MgO layer stacks at room temperature. Magneto-optical Kerr effect (MOKE) measurements also allow tracking skyrmion formation during external magnetic field sweeps to confirm their occurrence when measuring the topological Hall effect. We verify this outcome by comparing the results with thermal and electrical transport measurements from which we calculate the overall topological quantity that gives rise to the topological Nernst and Hall effect, respectively.

\section{Introduction}
\begin{figure}[ht]
\begin{center}
\begin{subfigure}{0.49\textwidth}
  \centering
  \includegraphics[width=\textwidth]{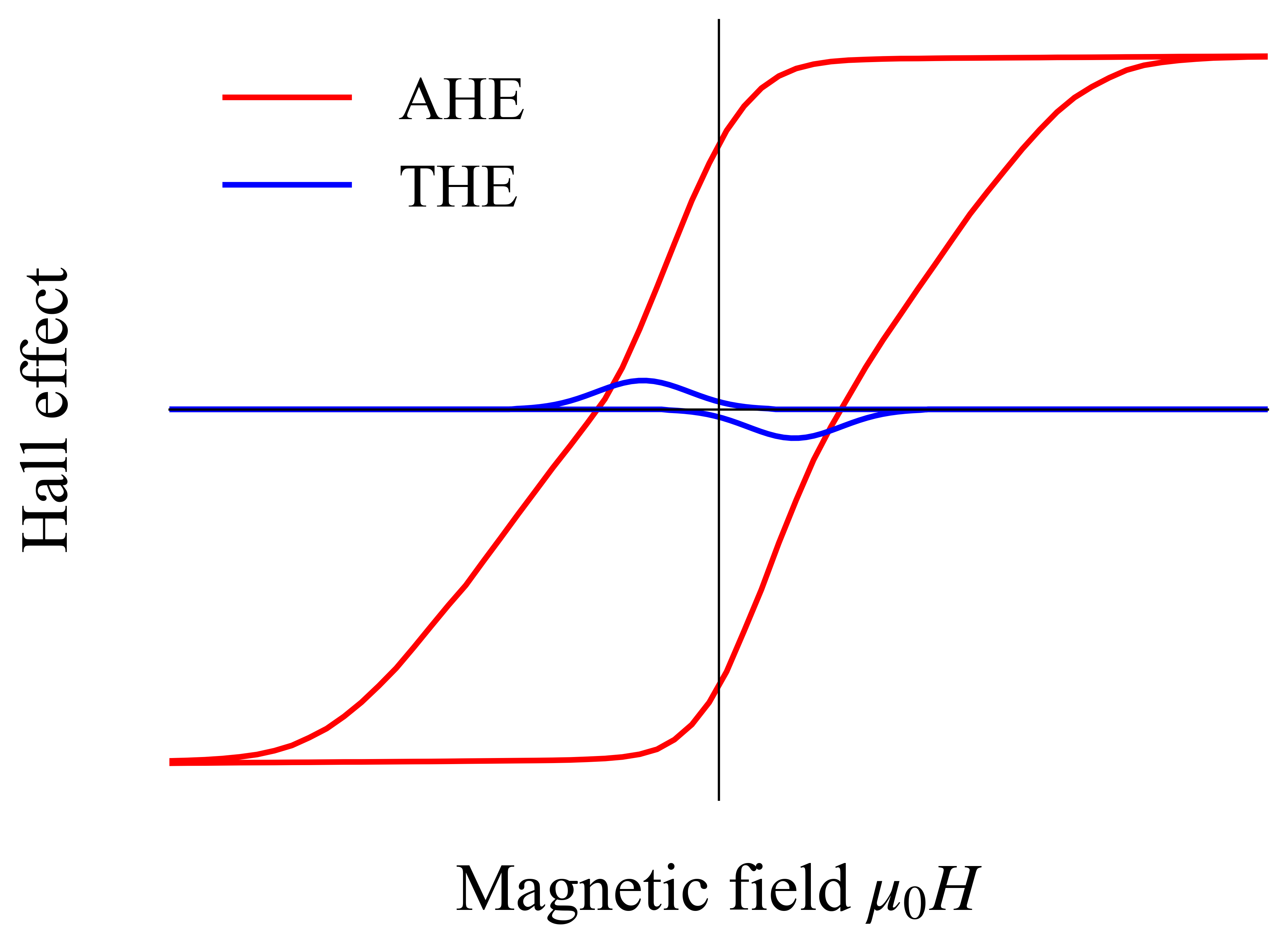}
  \caption{AHE and THE separated}
  \label{AHEuTHEPlot}
\end{subfigure}
\hfill
\begin{subfigure}{0.49\textwidth}
  \centering
  \includegraphics[width=\textwidth]{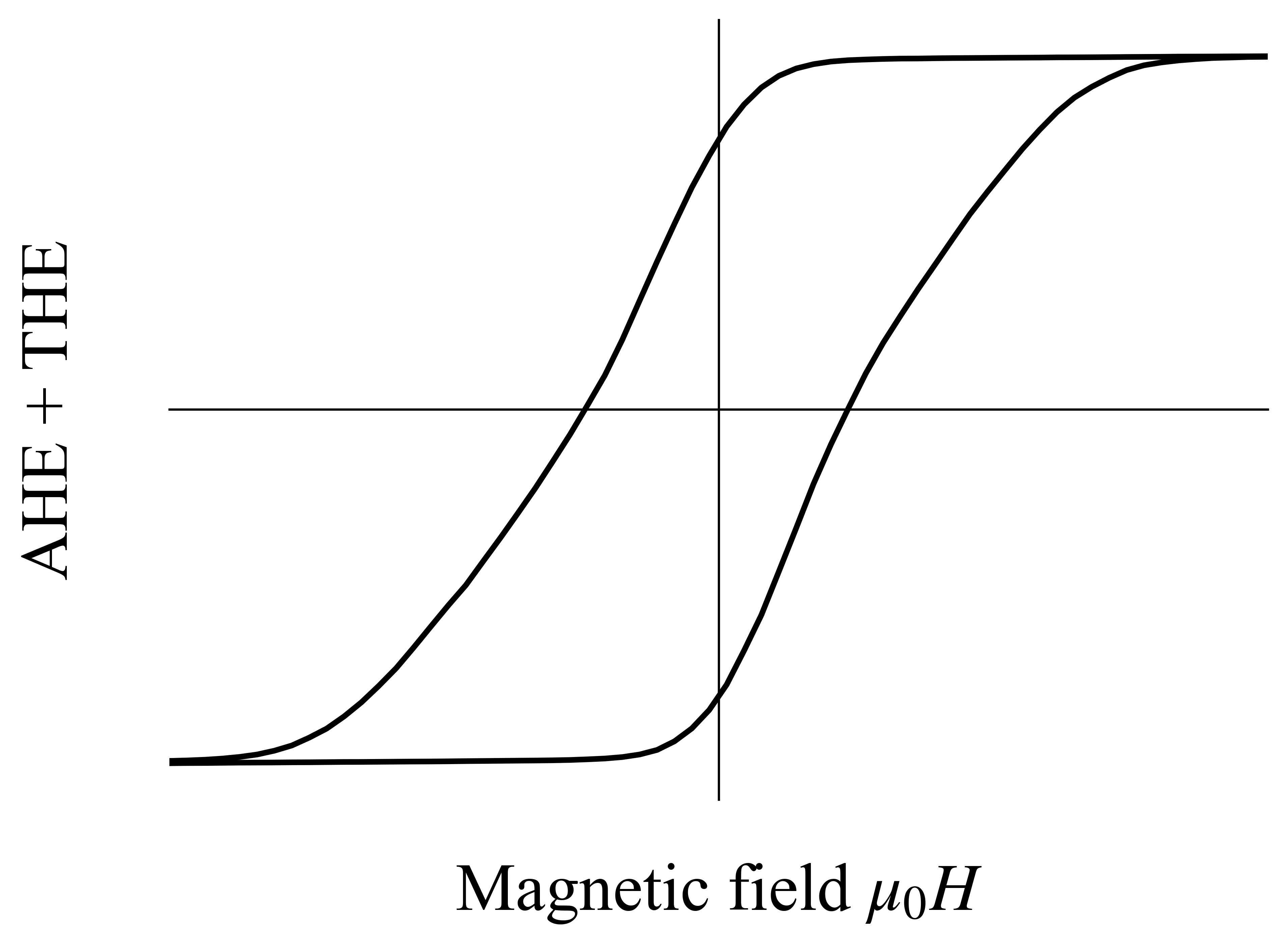}
  \caption{Sum of AHE and THE}
  \label{SumAHEuTHEPlot}
\end{subfigure}
\begin{subfigure}{0.49\textwidth}
  \centering
  \includegraphics[width=\textwidth]{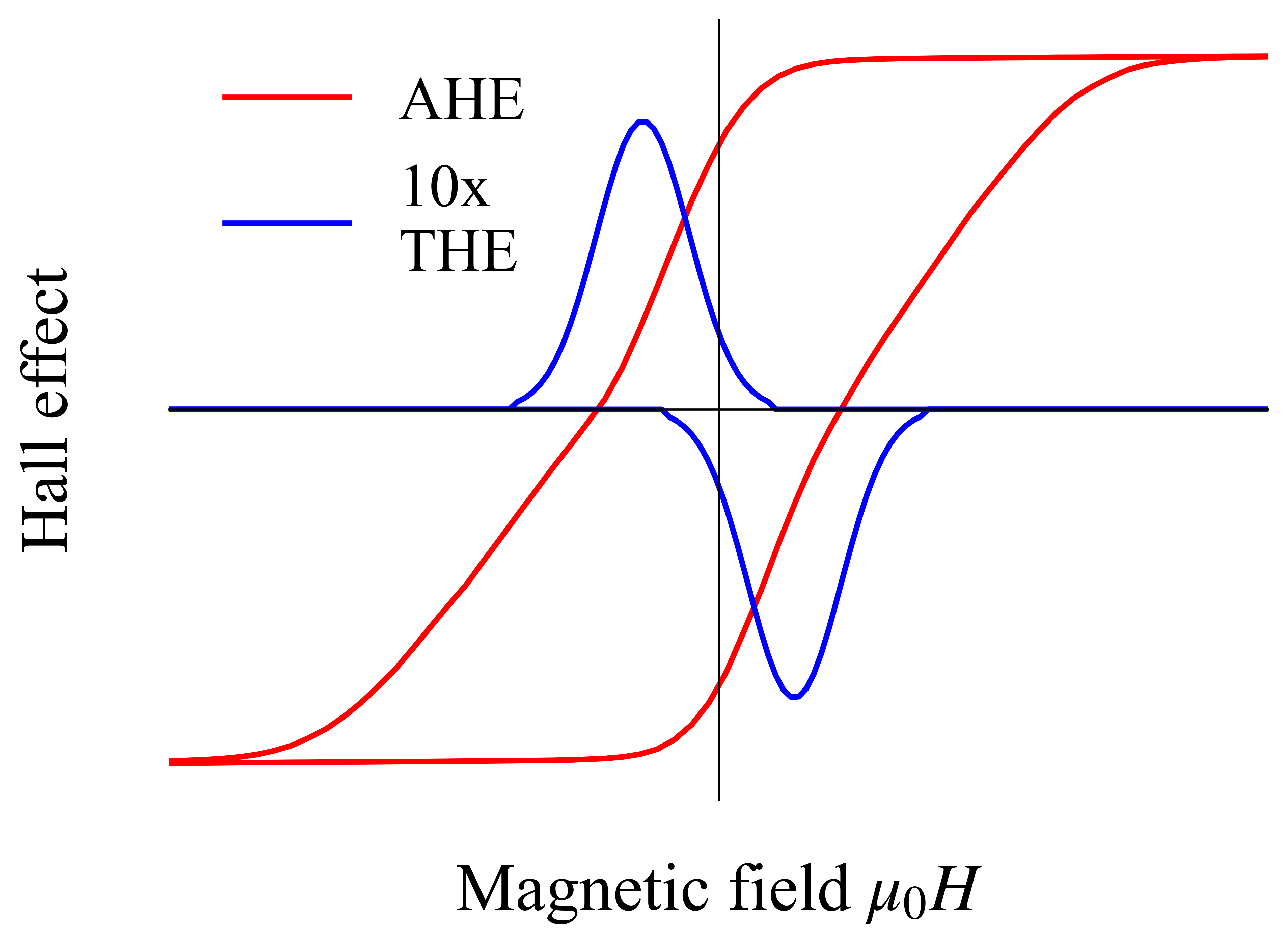}
  \caption{AHE and $10\mathrm{x}$ THE separated}
  \label{AHEu10THEPlot}
\end{subfigure}
\hfill
\begin{subfigure}{0.49\textwidth}
  \centering
  \includegraphics[width=\textwidth]{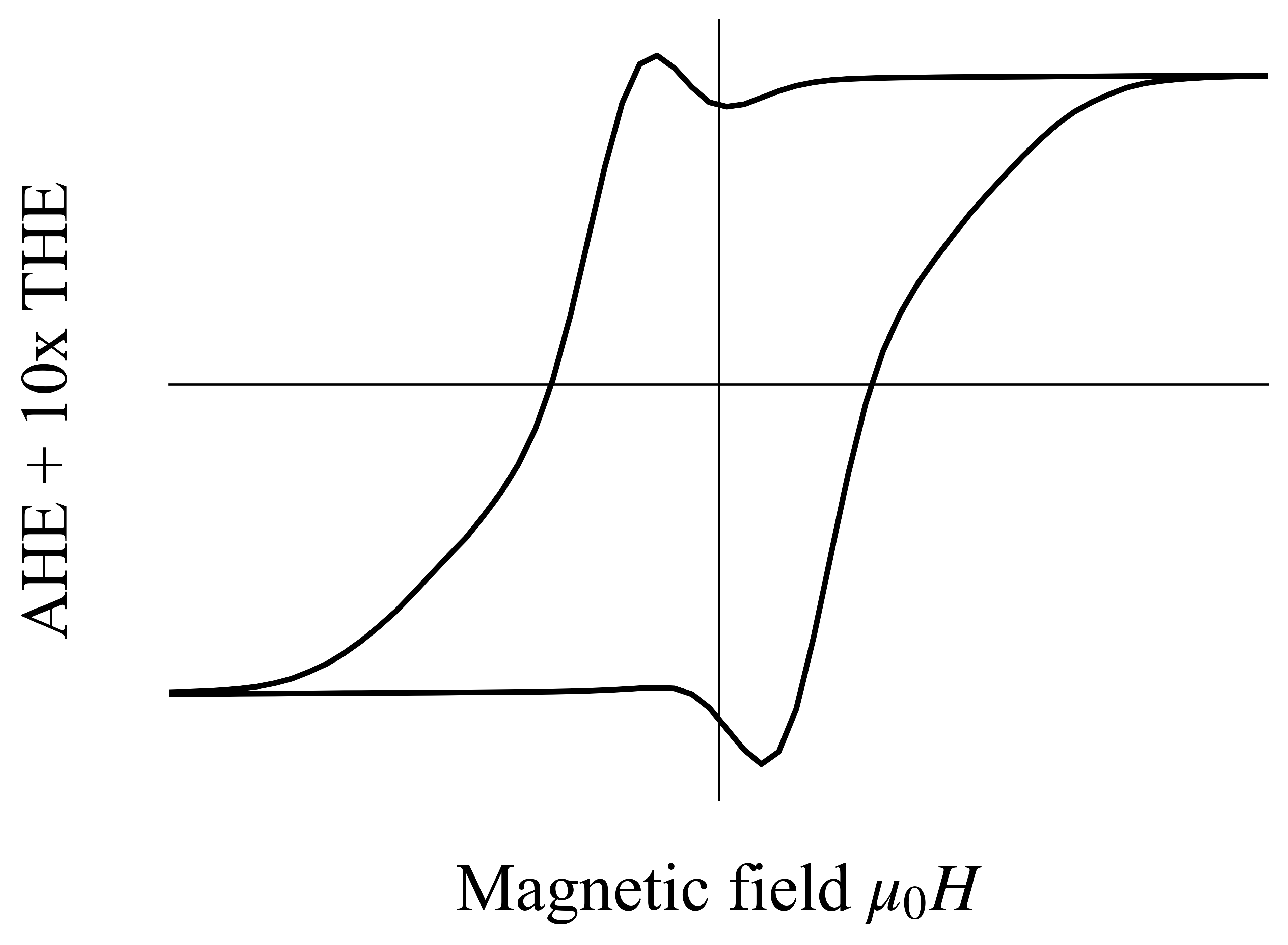}
  \caption{Sum of AHE and $10\mathrm{x}$ THE}
  \label{SumAHEu10THEPlot}
\end{subfigure}
\begin{subfigure}{0.49\textwidth}
  \centering
  \includegraphics[width=\textwidth]{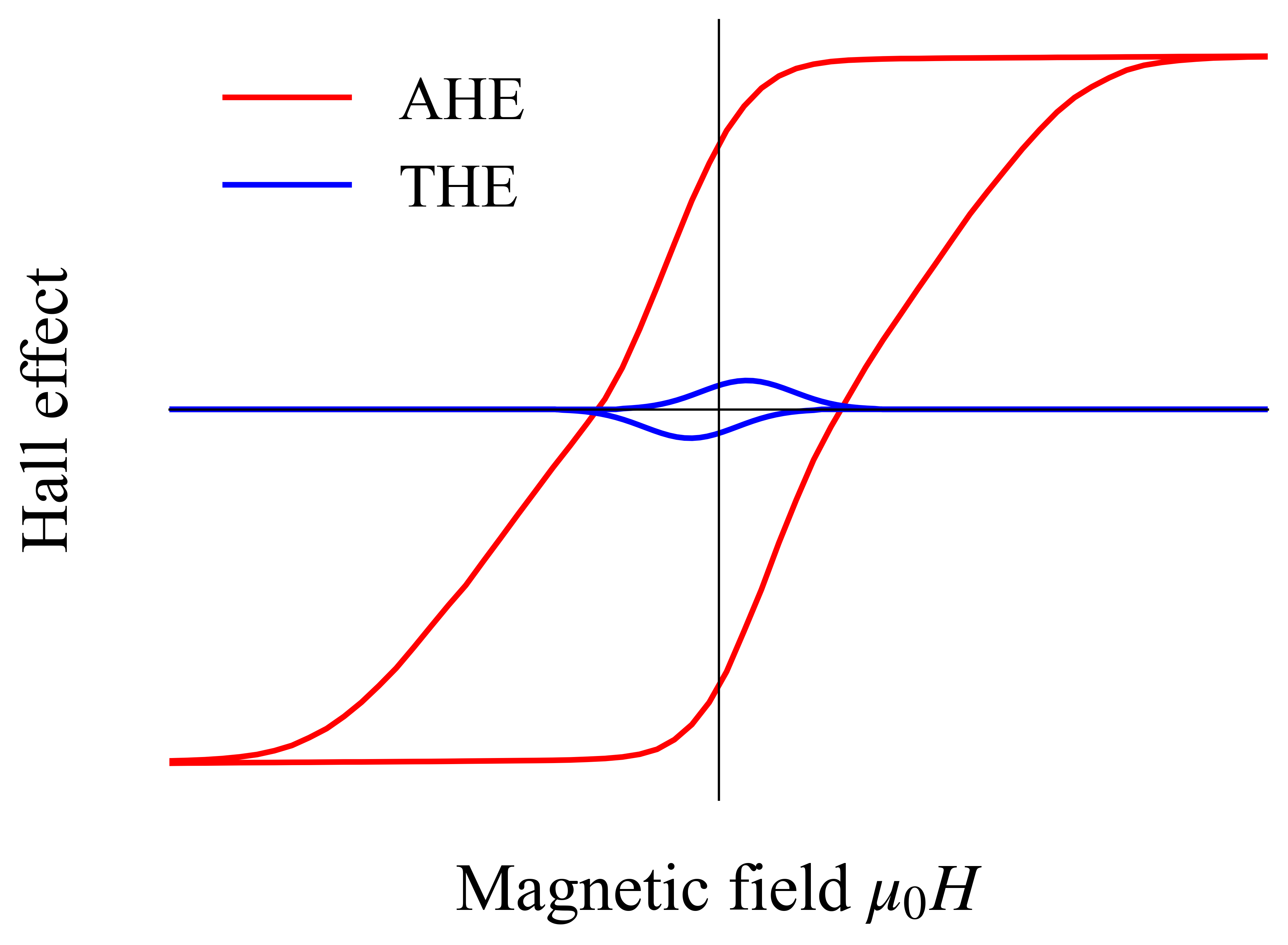}
  \caption{AHE and shifted THE separated}
  \label{AHEuShiftedTHEPlot}
\end{subfigure}
\hfill
\begin{subfigure}{0.49\textwidth}
  \centering
  \includegraphics[width=\textwidth]{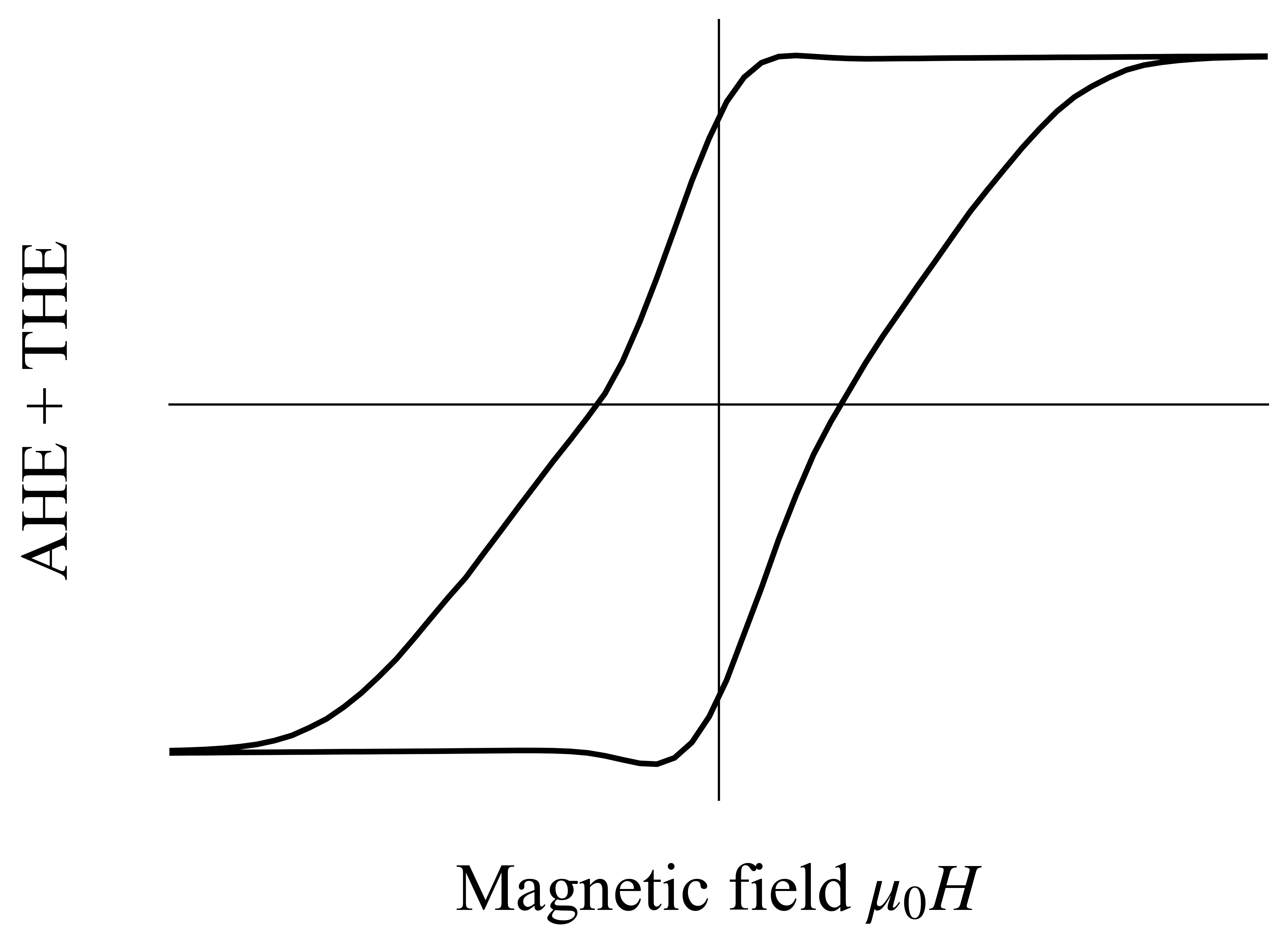}
  \caption{Sum of AHE and shifted THE}
  \label{SumAHEuShiftedTHEPlot}
\end{subfigure}
\caption{Schematic AHE and THE hysteresis loops based on measurement data to reveal the different contributions for varied relative strength. In (a,b) the THE peak height is $249(18)\,$\si{\pico\ohm\meter}, around $8.2\,\%$ of the saturation magnetization like measured for Ta/CoFeB/MgO, while for (c,d) the THE is multiplied by $10$ and for (e,f) shifted slightly. In Hall effect measurements of (b), the THE signal is hidden at the hysteresis loop edges, where skyrmions get nucleated and becomes clearly visible for higher THE peaks like in (d) or barely visible if it only gets shifted to a different region of the hysteresis loop like in (f).}
\label{SchematischeTHEAHEPlots}
\end{center}
\end{figure}

\begin{figure}[ht]
\begin{center}
\includegraphics[width=0.9\textwidth]{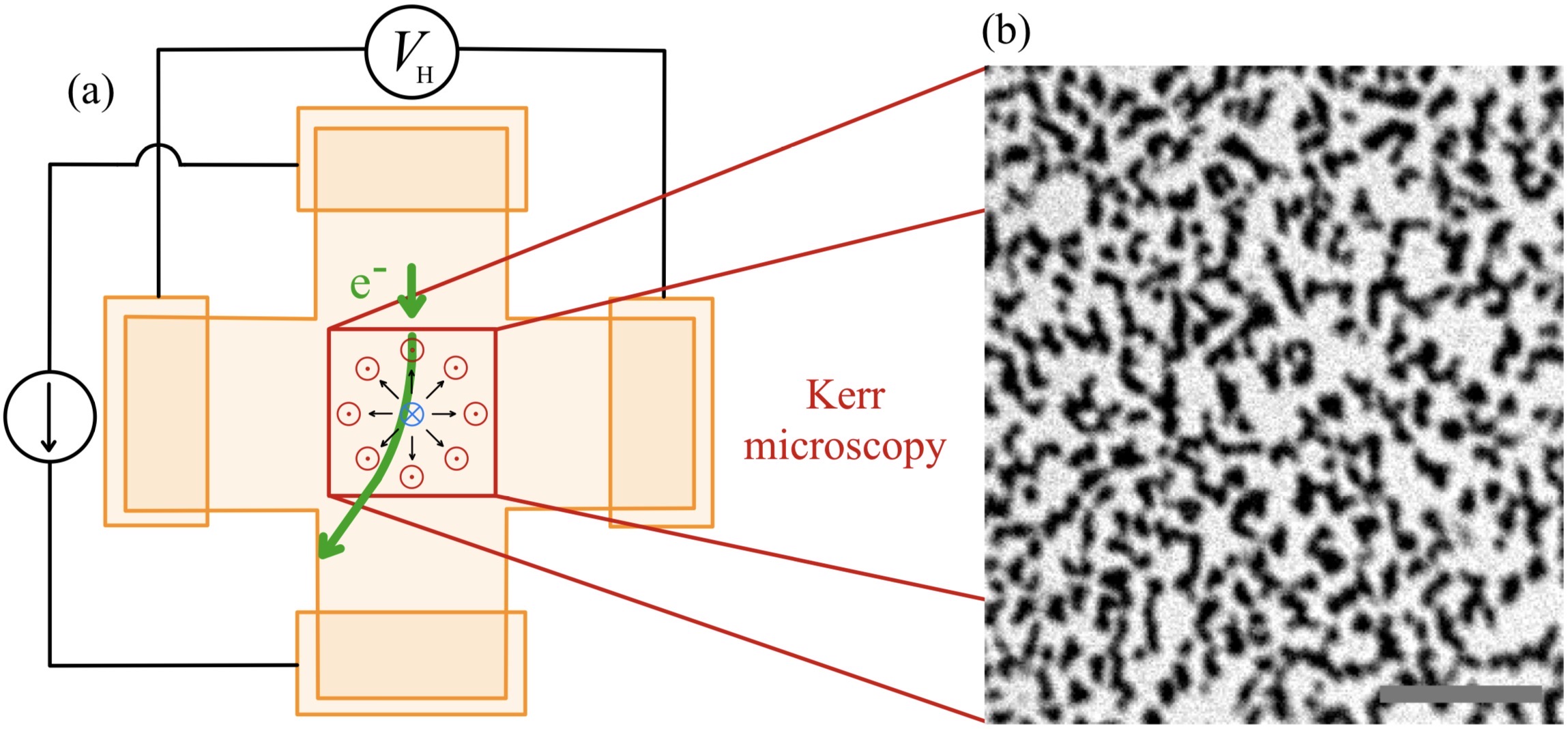}
\caption{(a) Schematic of a topological Hall effect (THE) measurement setup showing a Néel skyrmion with the winding number $Q=-1$ as an example centered inside the Hall cross. Electrons flowing through skyrmions are deflected by the magnetic field generated by the skyrmion spin structure. This electron deflection (THE) is detected in a Hall-cross configuration. This configuration requires separating the dominating anomalous Hall effect contribution, always present in the measurement signal, e.g. through magnetization measurements with Kerr microscopy (b). The image (b), also found in \autoref{KerrBildSkyrmionenB}, was acquired at $-0.28(5)\,$\si{\milli\tesla} and its scale bar corresponds to $5\,$\si{\micro\meter}. During measurements only out-of-plane magnetic fields are applied. The selection sketched in (a) is based on theory from \cite{Braun2012,Karnieli2021,Palotas2018}.}
\label{THEAufbauSchematischDraw}
\end{center}
\end{figure}

Magnetic skyrmions are two-dimensional topologically protected quasi particles consisting of a vortex-like spin structure. Their center and surrounding magnetization point in opposite out-of-plane directions. Important for their topology is the domain wall region, which is of Néel type for Ta/CoFeB/MgO single-layer films \cite{EverschorSitte2018,Qin2018,Yu2016} and does not allow a change of their topological charge $Q$ through continuous deformation for skyrmions. The latter criterion stabilizes the spin structure \cite{Tokura2020}. This specific spin structure gives rise to the topological Hall effect (THE) \cite{Neubauer2009,Bruno2004} and conversely, its observation indicates the presence of topologically protected spin textures. The THE is caused by non-coplanar magnetic textures such as skyrmions \cite{Leroux2018}, chiral bubbles, or certain domain wall configurations, which generate an emergent magnetic field that deflects moving electrons. In the adiabatic picture, electron spins follow the local magnetization and accumulate a real-space Berry phase, producing an extra transverse Hall signal beyond the ordinary and anomalous Hall effects \cite{Verma2022}. Imaging techniques such as magnetic force microscopy (MFM) \cite{Maccariello2018,Casiraghi2019}, Lorentz transmission electron microscopy (LTEM) \cite{Tang2019}, and Kerr microscopy \cite{McCord2025} provide valuable real-space information on magnetic domain structures, but do not verify their topological character. Therefore, measuring the topological Hall effect provides a complementary transport signature of nontrivial spin textures. While Kerr microscopy shows when magnetic particles or domains appear, the THE demonstrates that these textures also generate an emergent-field contribution to charge transport.\\
Various other measurements demonstrate the existence of the THE at low temperatures for bulk \cite{Neubauer2009,Kurumaji2019} and thin layer \cite{Schlitz2019,Li2013,Kimbell2022} materials, near room temperature in bulk materials \cite{Leroux2018} and at room temperature for thin layer materials \cite{Liu2022,A_Lone2024}. However, results showing the THE in single-layer magnetic films, like Ta/CoFeB/MgO at room temperature, have not been reported so far and this finding supports the topological nature of the magnetic structures. In this material, the weak THE is concealed in the Hall-effect hysteresis loops, as shown schematically in Figures \ref{AHEuTHEPlot} and \subref{SumAHEuTHEPlot}. The small THE peaks appear at the hysteresis edges, where the magnetic signal changes rapidly. Measuring the THE is challenging because it requires separation from the anomalous Hall effect (AHE). Larger or shifted peaks, found in other materials, improve the visibility of the THE. For higher THE peaks, like in \autoref{AHEu10THEPlot} and \subref{SumAHEu10THEPlot}, the THE would be directly visible in the hysteresis loop shape of the total Hall effect measurement. On the other hand, small shifts of the THE peaks, like in \autoref{AHEuShiftedTHEPlot} and \subref{SumAHEuShiftedTHEPlot}, just slightly improve the visibility of the THE.\\
In this publication, we demonstrate how to extract the THE from concurrent electrical Hall effect measurements and Kerr microscope images of the surface magnetization, recorded while cycling the applied magnetic field with the setup of \autoref{THEAufbauSchematischDraw}. We observe point-like structures in the Kerr images and peaks representing the THE for the same applied fields. This simultaneous occurrence indicates that these structures are topologically protected objects, most likely skyrmions. This THE-measurement method does not require any deployment of high resolution imaging techniques and can be employed even for single nanometer thick ferromagnetic films. It could therefore complement high temporal resolution techniques such as time resolved magneto-optical Kerr effect (TR-MOKE) measurements \cite{Kalin2024} by providing additional information on the topology of the investigated magnetic structures.\\
We verify the outcome by additional measurements, extracting the THE from the topological Nernst effect (TNE), and Hall effect measurements performed on specially developed sample structures. The topological quantity is determined as described in reference \cite{Schlitz2019}. Spin textures, e.g., skyrmions, cause both effects, the THE and the TNE, so that the combination of both experiments confirms the existence of the weak THE as a material property of single-layer Ta/CoFeB/MgO films at room temperature.

\clearpage

\clearpage

\section{Theory}

\subsection{Extracting the topological Hall effect}
In conducting ferromagnetic thin-film heterostructures with interfacial Dzyaloshinskii–Moriya interaction \cite{Khan2016}, such as Ta/CoFeB/MgO, skyrmion-like spin textures can form. In such systems, the measured Hall resistivity $\rho^{\mathrm{Hall}}$ generally consists of ordinary, anomalous, and topological contributions:
\begin{equation}
\rho^\mathrm{Hall} = \rho^\mathrm{OHE} + \rho^\mathrm{AHE} + \rho^\mathrm{THE}
\label{GesHallFormel}
\end{equation}
$\rho^\mathrm{OHE}=R_0\mu_0H$ is the ordinary Hall effect (OHE), $\rho^\mathrm{AHE}=R_\mathrm{S}M$ is the AHE and $\rho^\mathrm{THE}$ is the THE \cite{Leroux2018}. The OHE exhibits a linear dependence on the magnetic field $\mu_0H$ with the proportionality constant given by the OHE constant $R_0$. Here $\mu_0$ is the magnetic constant. Analogously, the AHE $\rho^\mathrm{AHE}$ contribution is proportional to the magnetization $M$, with the proportionality given by the anomalous Hall coefficient $R_\mathrm{S}$. In contrast, the THE is caused by the spin texture of the skyrmions \cite{Leroux2018}.\\
In a simplified model introduced in references \cite{Neubauer2009,Bruno2004,Machida2007}, the THE is described by:
\begin{equation}
\rho^\mathrm{THE} \approx P R_0 B^z_\mathrm{eff},
\label{THEFormelTheorie}
\end{equation}
with $P$ the local spin polarization of the conduction electrons and an effective field $B^\mu_\mathrm{eff} = \Phi_0 \Phi^\mu$. This consists of the flux quantum for a single electron $\Phi_0=h/e$ and the flux density calculated by the skyrmion density:
\begin{equation}
\Phi^\mu = \frac{1}{8 \pi}\epsilon_{\mu \nu \lambda} \hat{m}\cdot(\partial_\nu \hat{m}\times\partial_\lambda\hat{m}).
\end{equation}
Here $\epsilon_{\mu \nu \lambda}$ is the Levi-Civita symbol and $\hat{m}=\vec{M}/|M|$ is the normalized magnetization.\\
In our material, the THE is much smaller than the AHE and we separate both quantities by calculating the sample magnetization from the brightness of the magneto-optic Kerr effect (MOKE) images.\\
The interpretation of \autoref{THEFormelTheorie} is most straightforward for isolated skyrmions or dilute skyrmion ensembles, where the emergent magnetic field can be related to the sum of individual topological spin textures. In more complex multidomain states however, the relation between the measured Hall signal and the number of skyrmions is less direct. Labyrinthine domains, stripe-domain end caps, and topological defects can carry spatially distributed topological charge densities and influence the transition between multidomain states and skyrmions through domain wall motion and the generation or deletion of topological defects, as shown for Ta/CoFeB/MgO multilayers \cite{Wang2019}. Moreover, recent random-resistor-network calculations show that nonmonotonic Hall features resembling a THE can also arise from AHE contributions combined with domain wall scattering in disordered multidomain ferromagnets, even without a finite topological charge \cite{Sabri2025}. Therefore, in dense or merging domain states, a THE-like signal cannot be interpreted solely in terms of the number of skyrmions, but requires comparison with the magnetic domain configuration and careful separation from AHE-related contributions.

\subsection{Nernst effect}
The Nernst effect functions similarly to the Hall effect, but instead of applying a potential difference and driving a current perpendicular to the applied magnetic field, it arises from a temperature gradient in this direction. The temperature gradient generates a longitudinal thermoelectric charge-carrier flow, the Seebeck effect. Simultaneous sample magnetization or applied fields pointing perpendicular to the temperature gradient deflect charge carriers perpendicular to both the field or magnetization and the thermoelectric flow, causing the Nernst effect.\\
The complete measured Nernst effect $N_\mathrm{xy}$ signal composes of the following contributions:
\begin{equation}
N^\mathrm{Nernst}_\mathrm{xy} = N^\mathrm{ONE}_\mathrm{xy} + N^\mathrm{ANE}_\mathrm{xy} + N^\mathrm{TNE}_\mathrm{xy},
\label{GesNernstFormel}
\end{equation}
which is analogous to the Hall effect in \autoref{GesHallFormel}. Here $N^\mathrm{ONE}_\mathrm{xy}=N_\mathrm{O} (\mu_0 \vec{H}\times \Delta T)$ is the ordinary Nernst effect (ONE) with the ordinary Nernst coefficient $N_\mathrm{O}$, the temperature gradient $\Delta T$ and the magnetic field vector $\vec{M}$ \cite{Mizuguchi2019}, $N^\mathrm{ANE}_\mathrm{xy}=N_\mathrm{A}(\vec{M}\times \Delta T)$ is the anomalous Nernst effect (ANE) with the anomalous Nernst coefficient $N_\mathrm{A}$ and magnetization $\vec{M}$ \cite{Gamino2021} and $N^\mathrm{TNE}_\mathrm{xy}$ is the topological Nernst effect (TNE), is caused by the spin structure of the skyrmions analogous to the THE, described in more detail in references \cite{Madhogaria2023,Asaba2021,Hirschberger2020}.\\
The similarity between Equations \ref{GesHallFormel} and \ref{GesNernstFormel} allows to calculate the topological quantity introduced in reference \cite{Schlitz2019} from the difference of both signals. After normalizing both signals to their saturation magnetization, the topological quantity:
\begin{equation}
N^\mathrm{TQ} = \rho^\mathrm{Hall}_\mathrm{norm.}-N^\mathrm{Nernst}_\mathrm{xy, norm.} = \rho^\mathrm{THE}_\mathrm{norm.}-N^\mathrm{TNE}_\mathrm{xy, norm.}
\label{TopologicalQuantityFormel}
\end{equation}
isolates the topological contributions $\rho^\mathrm{THE}_\mathrm{norm.}$ and $N^\mathrm{TNE}_\mathrm{xy, norm.}$ by eliminating all normal and anomalous parts. Both remaining components, the THE and TNE, can have opposite signs and magnitudes and therefore remain in $N^\mathrm{TQ}$ rather than canceling each other out.\\
Instead of using the magnetization information to extract the THE from electrical Hall effect measurements with MOKE, \autoref{TopologicalQuantityFormel} uses the Nernst effect to calculate the $N^\mathrm{TQ}$ that contains a combination of THE and TNE.\\

\clearpage
\section{Methods}
\subsection{Layer stack deposition for skyrmion generation}
\begin{figure}[ht]
\begin{center}
\begin{subfigure}{0.51\textwidth}
  \centering
  \includegraphics[width=\textwidth]{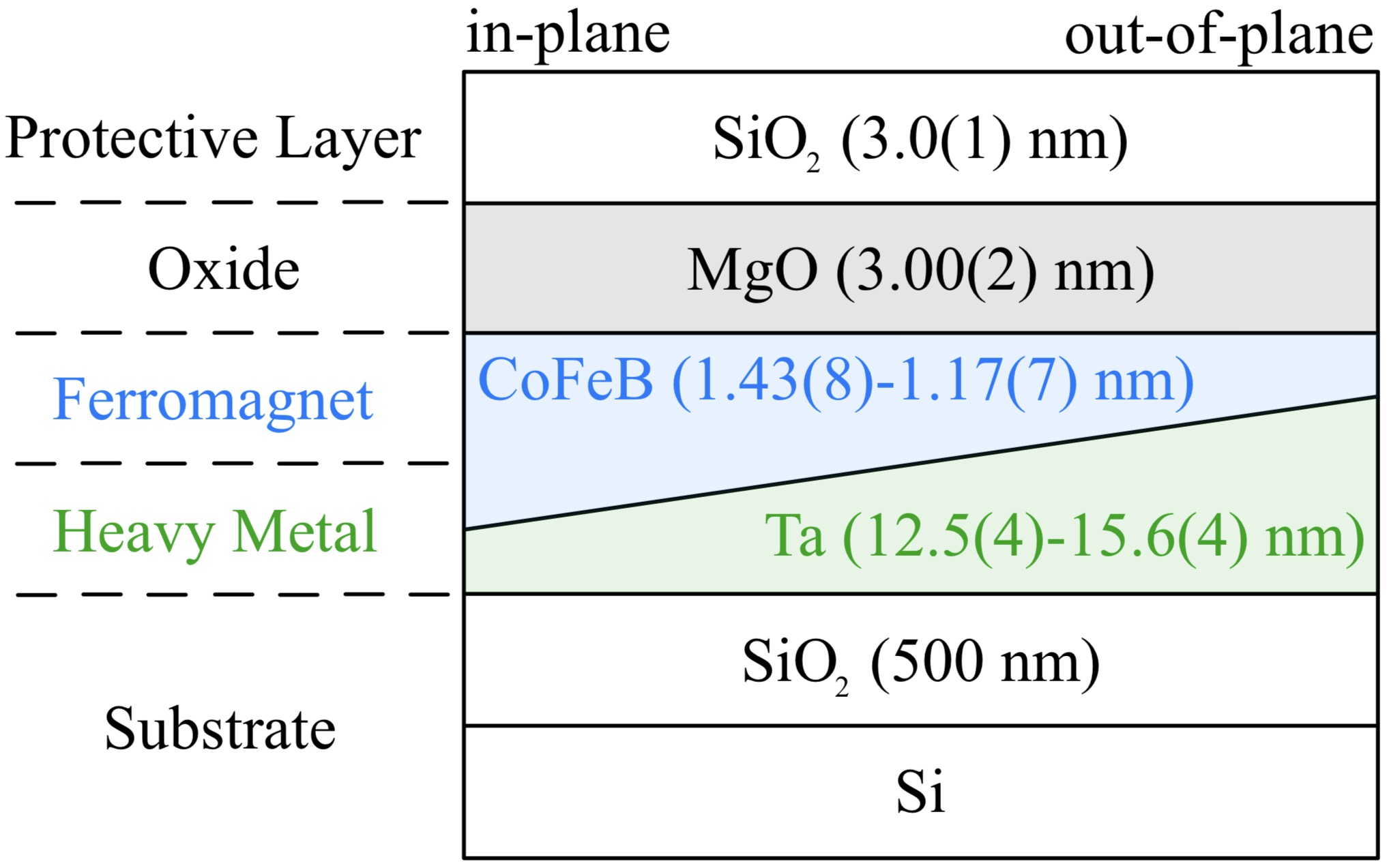}
  \caption{Skyrmion sample layer stack}
  \label{LayerStackBild}
\end{subfigure}
\hfill
\begin{subfigure}{0.45\textwidth}
  \centering
  \includegraphics[width=\textwidth]{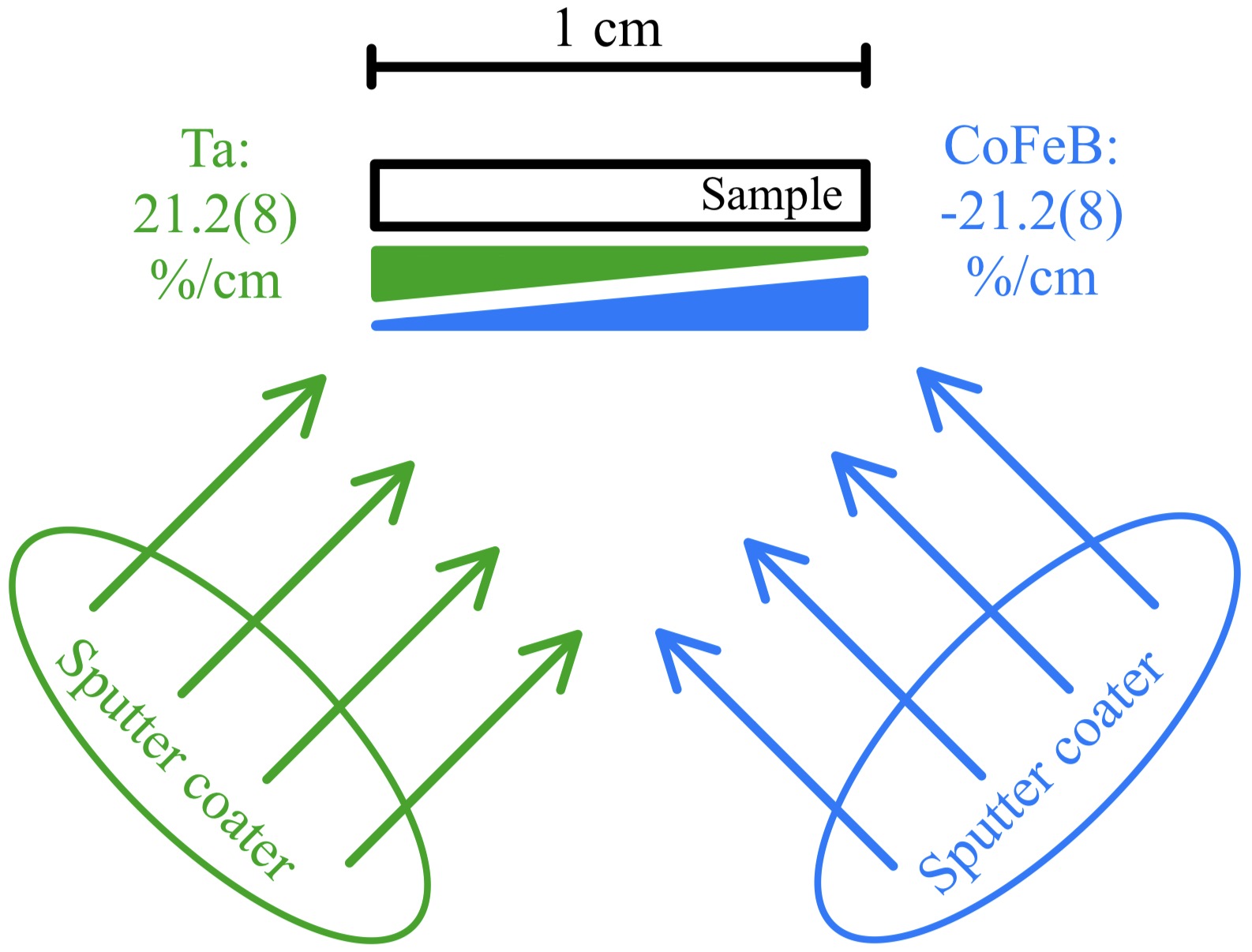}
  \caption{Sample with sputter geometry}
  \label{CoFeBGradients}
\end{subfigure}
\caption{(a) Layer stack for skyrmion creation using sputter coating for Ta and CoFeB with opposing thickness gradients of $\pm 21.2(8)\, \mathrm{\%/}$\si{\centi\meter} and e-beam evaporation for MgO and SiO$\mathrm{_2}$. The layer thicknesses are measured with quartz crystal microbalances and are calculated for both sample edges along the thickness gradient using reference \cite{zbarsky2014spindynamik}. The CoFeB thickness gradient generates a landscape of regions with magnetic out-of-plane and in-plane anisotropies. At the transition region between both anisotropy types, skyrmions can be created. These parameters are valid for the MOKE-Hall sample and the Nernst-Hall sample differs by using CoFeB thicknesses of $1.50(8)\,$\si{\nano\meter} to $1.23(7)\,$\si{\nano\meter}. (b) Schematic of the sample and sputter deposition geometry, introducing thickness gradients in the Ta and CoFeB layers.}
\label{LayerStackUndCoFeBGradients}
\end{center}
\end{figure}

\autoref{LayerStackBild} depicts the layer stack deposited on a thermally oxidized mono crystalline substrate by magnetron sputtering (Ta and CoFeB) and e-beam evaporation (MgO and SiO$_\mathrm{2}$). A quartz crystal microbalance in combination with a shutter enable precise layer thickness control. A gradient of $21.2(8)\, \mathrm{\%/}$\si{\centi\meter} in Ta and $-21.2(8)\, \mathrm{\%/}$\si{\centi\meter} in CoFeB with the geometry depicted in \autoref{CoFeBGradients} is achieved with tilting the sputter sources relative to the sample. In this configuration, especially the gradient achieved in the CoFeB layer allows for a fine control over the effective out-of-plane magnetic anisotropy across the sample to get the right conditions for skyrmion creation in the transition area between out-of-plane and in-plane anisotropy. The skyrmion size and density are highly sensitive to the CoFeB layer thickness, like investigated in reference \cite{Denker2023}. After deposition, the samples are annealed for $1\,$\si{\hour} at $473\,$\si{\kelvin} at a base pressure around $10^{-4}\,$\si{\pascal}, adjusting multiple material properties to favor skyrmion formation \cite{Khan2016,Wang2006,Zhu2012}.

\newpage
\subsection{Micro structures and electrical circuits}
\begin{figure}[ht]
\begin{center}
\begin{subfigure}[b]{0.38\textwidth}
  \centering
  \includegraphics[width=\textwidth]{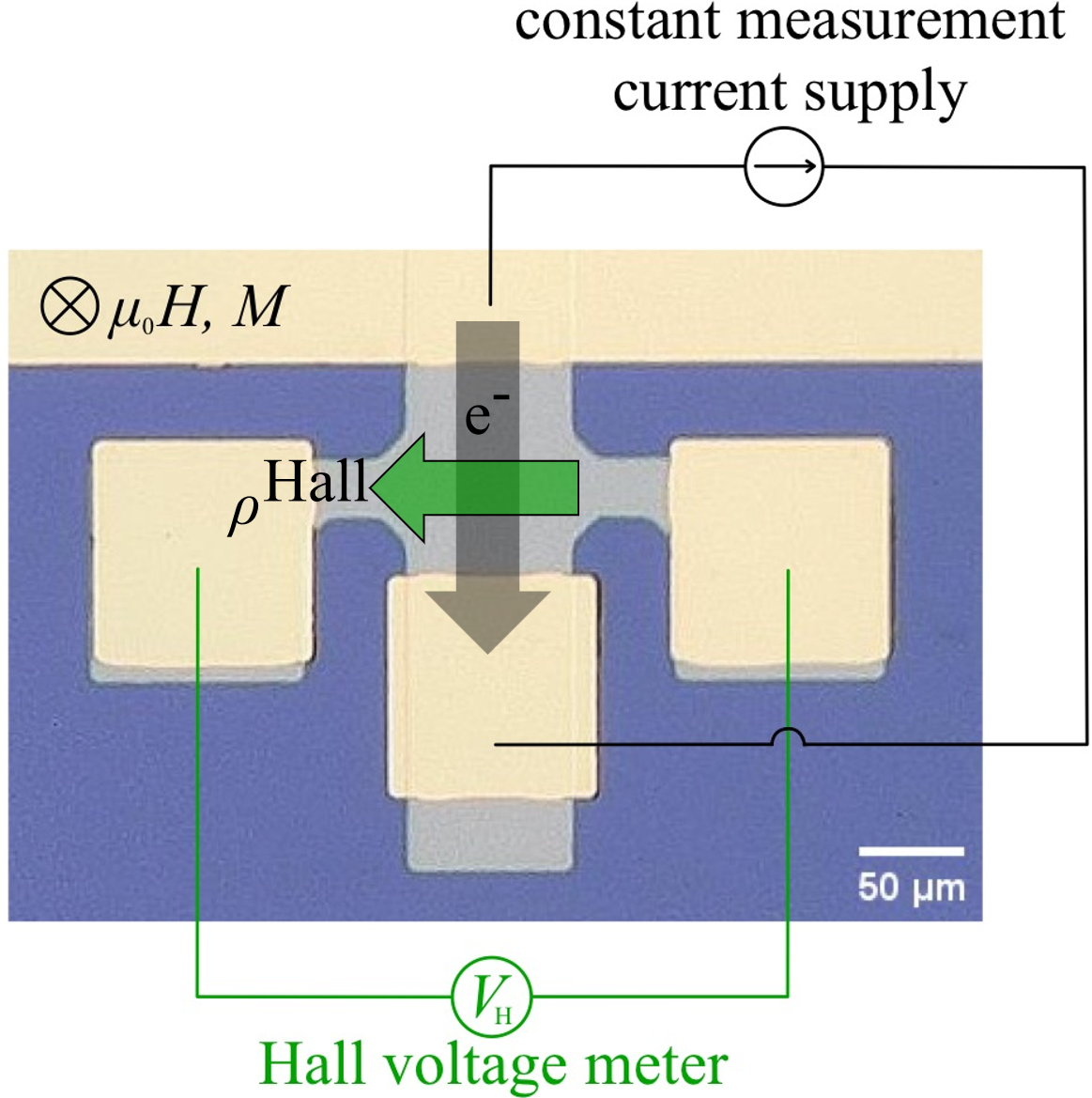}
  \caption{Hall cross for MOKE}
  \label{SampleTHE}
\end{subfigure}
\hfill
\begin{subfigure}[b]{0.59\textwidth}
  \centering
  \raisebox{7mm}{
    \includegraphics[width=\textwidth]{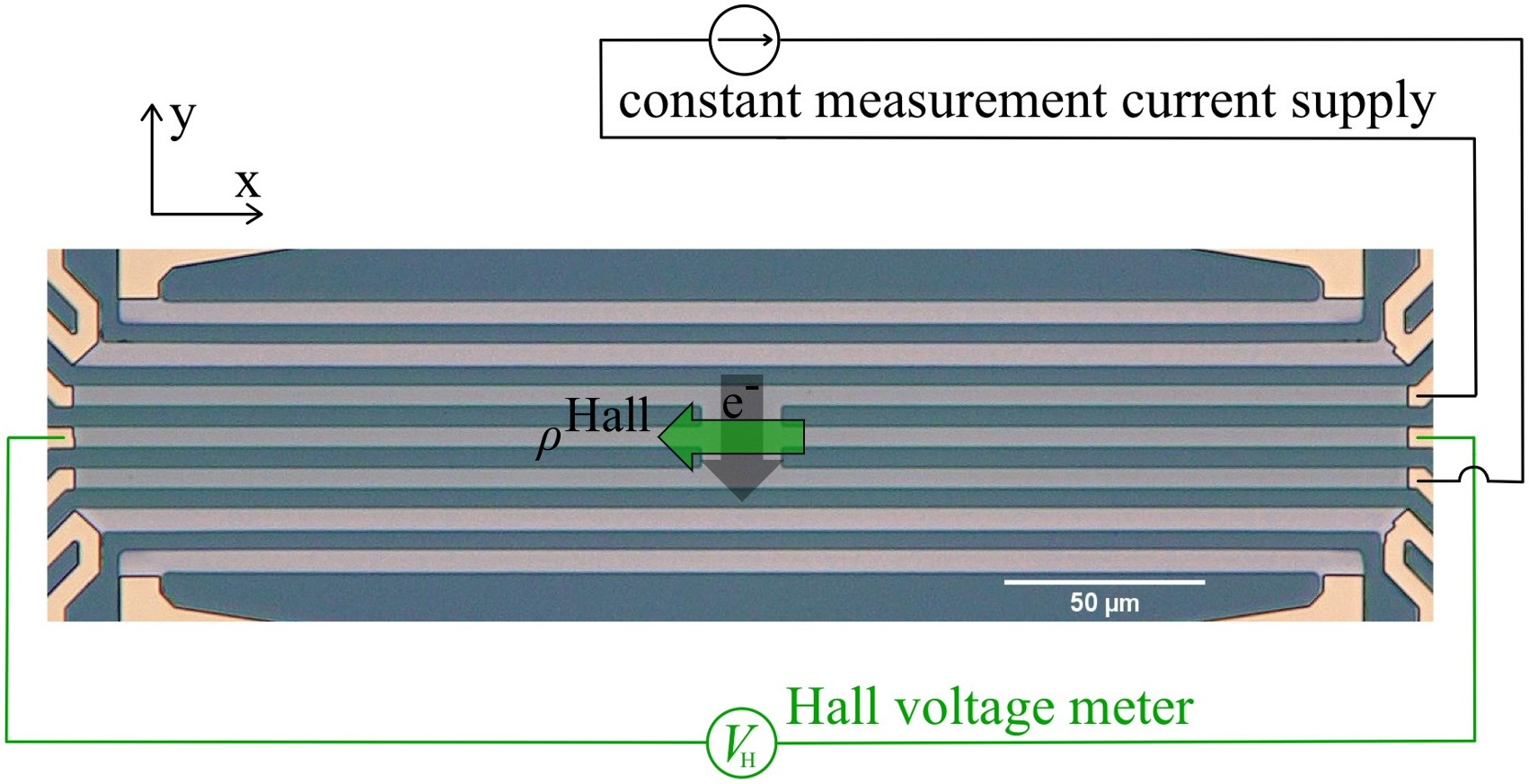}
  }  
  \caption{Hall effect measurement at thermal transport sample}
  \label{SampleStructureTNEHall}
\end{subfigure}
\begin{subfigure}[c]{0.59\textwidth}
  \centering
  \includegraphics[width=\textwidth]{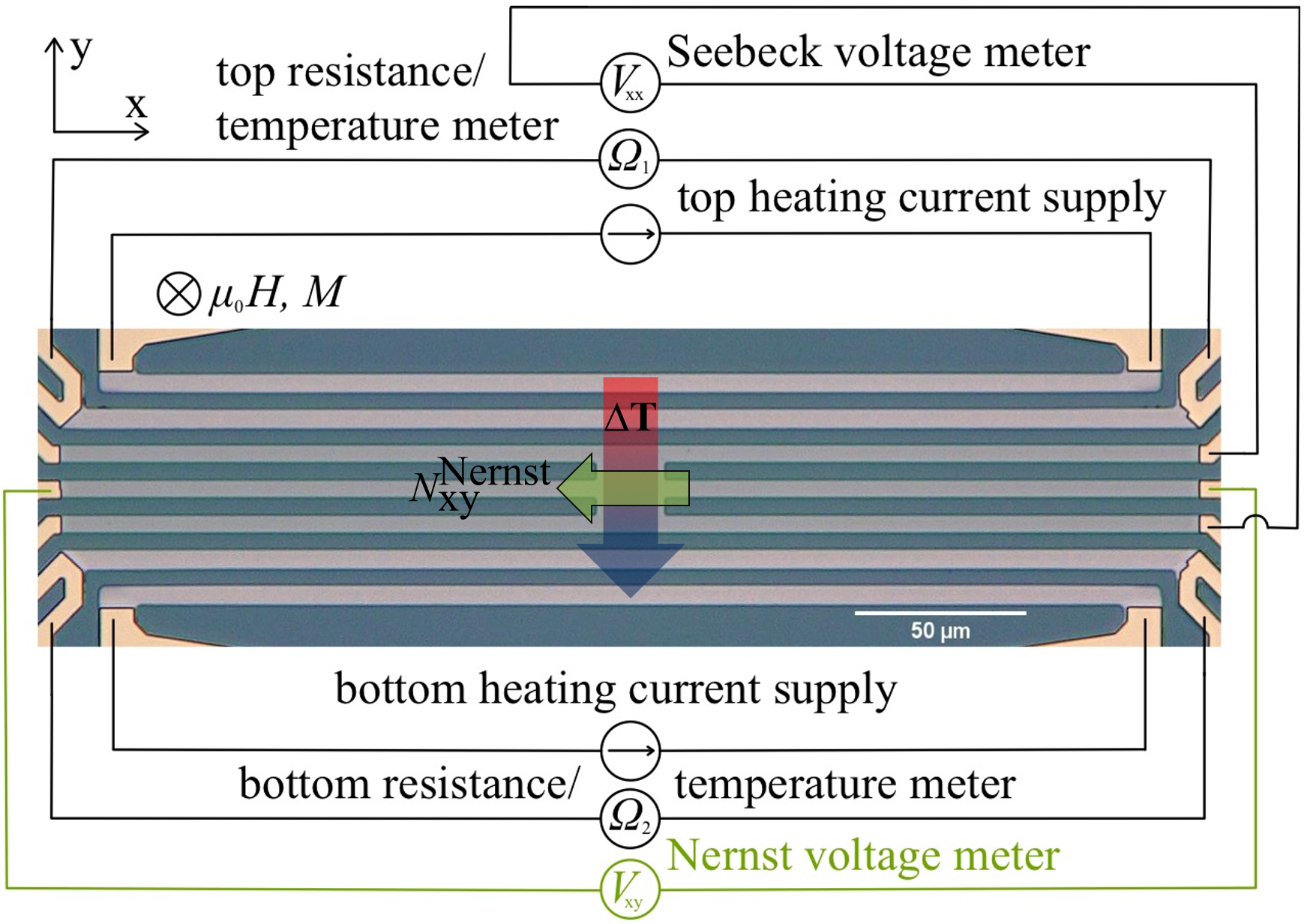}
  \caption{Nernst effect measurement at thermal transport sample}
  \label{SampleStructureTNENernst}
\end{subfigure}
\caption{Optical microscope images of the investigated devices. (a,b) Cross structures for Hall effect measurements with out-of-plane magnetic fields. (a) has a larger surface area, better suited for MOKE measurements and (b) is smaller allowing for larger temperature gradients when measuring the Nernst effect in (c). Both use constant measurement currents. 
(c) Micro structure and electrical measurement circuit for measuring the Nernst effect. Its structure is adapted from \cite{Schlitz2019} with minor modifications in the cross width. The sightly greenish shimmering cross in the center consist of the skyrmion material shown in \autoref{LayerStackBild}. Above and below it, are temperature sensors located, while heaters are placed at the outermost regions. These sensors and heaters consist of $5\,$\si{\micro\meter}-wide and $33\,$\si{\nano\meter}-thick chromium/platinum lines. While measuring, constant heating currents and out-of-plane magnetic fields are applied.}
\label{THEuTNESampleStructure}
\end{center}
\end{figure}

Figures \ref{SampleTHE} and \subref{SampleStructureTNEHall} depict the patterned layer stacks after the lift-off and argon ion milling process. The Hall cross shown in \autoref{SampleTHE} is patterned employing simple mask photolithography and was developed for the simultaneous Hall effect measurement and Kerr microscopy monitoring of the surface magnetization (MOKE-Hall sample). The thermal transport measurements require more precise features, and therefore, the patterns shown in \autoref{SampleStructureTNEHall}, integrating heater lines and thermometers into the Hall cross layout, are fabricated using e-beam lithography (Nernst-Hall sample). Both samples depict the wiring configuration for Hall measurements with an out-of-plane magnetization direction. All electrical signals were averaged from $32$ individual subsequently acquired measurement points unless otherwise stated.\\
Constant currents with $5.00(3)\,$\si{\milli\ampere} (\autoref{SampleTHE}) and $0.10000(4)\,$\si{\milli\ampere} (\autoref{SampleStructureTNEHall}) (including source accuracy), applied in the y-direction, generate Hall voltages in the x-direction. The Hall resistivity $\rho_\mathrm{Hall}$ follows from:
\begin{equation}
\rho_\mathrm{Hall}=\frac{V_\mathrm{H}\cdot d_\mathrm{CoFeB+Ta}}{I_\mathrm{Long}},
\label{HallResistivityFormel}
\end{equation}
where $V_\mathrm{H}$ is the Hall voltage, $I_\mathrm{Long}$ represents the longitudinal measurement current, and $d_\mathrm{CoFeB+Ta} = d_\mathrm{CoFeB} + d_\mathrm{Ta}$ is the thickness of the complete electrically conductive material \cite{gross2014festkorperphysik}. Both thicknesses are calculated employing the gradients and the specific measured layer thicknesses introduced in \autoref{LayerStackBild} for the investigated structure positions on the samples. This results in a thickness $d_\mathrm{MOKE,\, Hall,\, CoFeB+Ta} = 1.25(7)\, \si{\nano\meter} + 14.6(4)\, \si{\nano\meter} = 15.9(4)\, \si{\nano\meter}$ for the MOKE-Hall sample, and $d_\mathrm{Nernst,\, Hall,\, CoFeB+Ta} = 1.41(8)\, \si{\nano\meter} + 13.5(4)\, \si{\nano\meter} = 14.9(4)\, \si{\nano\meter}$ for the Nernst-Hall sample.\\
\autoref{SampleStructureTNENernst} shows the Nernst-Hall sample wired in the configuration for the Nernst effect measurement, adapted from reference \cite{Schlitz2019}. Applying a $10\,$\si{\milli\ampere} current through one of the two Joule heater lines creates a temperature gradient in the y-direction. Changing the heater line reverses the direction of the temperature gradient. The resulting temperature difference is detected by the two resistive temperature sensor lines, which are composed of $3\,$\si{\nano\meter} Cr and $30\,$\si{\nano\meter} Pt layers, where the Cr layer promotes the adhesion between Pt and the SiO$_2$ substrate surface. Its dimensions are chosen to approximate resistors with approximately $1\,$\si{\kilo\ohm} at room temperature, functionally similar to Pt1000 temperature sensors. All temperature sensors are measured with $100\,$\si{\micro\ampere} currents and are calibrated with a Pt1000 sensor using a heating plate.\\
The temperature gradient creates an electron movement towards the colder region in the y-direction, known as the Seebeck voltage. The out-of-plane magnetization deflects those electrons perpendicularly to the x-direction, generating the Nernst voltage. Before each Nernst effect measurement, the heaters are operated for 5 minutes to establish an equilibrium temperature gradient amounting to $0.0218(4)\,$\si{\kelvin/\micro\meter}. Like in reference \cite{Schlitz2019}, all Nernst measurements are performed applying both temperature gradient directions, and combining them into the Nernst effect:
\begin{equation}
N_{xy}=\frac{d_\mathrm{therm}}{l_\mathrm{H}} \cdot \frac{V_\mathrm{xy}^\mathrm{up}-V_\mathrm{xy}^\mathrm{down}}{\Delta T^\mathrm{up}-\Delta T^\mathrm{down}},
\end{equation}
and thus, removing all parasitic thermoelectric contributions. Here, $d_\mathrm{therm}=40(1)\,$\si{\micro\meter} represents the distance between both parallel aligned Pt thermal couples, $l_\mathrm{H}=309(2)\,$\si{\micro\meter} is the length of the heating elements, and $V^\mathrm{up/down}_\mathrm{xy}$ is the measured Nernst voltage for the up or down direction of the temperature gradient $\Delta T^\mathrm{up/down}$.\\
The Hall resistivity of the Hall-MOKE sample has a relative systematic error of $4.5\,$\%, while the Hall-Nernst sample has a relative systematic error of $1.8\,$\% for the Hall and $8.4\,$\% for the Nernst measurement.

\clearpage
\subsection{Magneto optical Kerr effect microscopy}
\label{MOKETheory}
A Kerr microscope \cite{McCord2015} uses the MOKE to image polarization changes of the reflected light due to magnetization changes of reflective magnetic surfaces and measures relative magnetization direction changes as intensity changes. The microscope resolution is limited for the optical setup by the Rayleigh criterion of around $300\,$\si{\nano\meter} at a wavelength $\lambda=450\,$\si{\nano\meter} and a numerical aperture $\mathrm{NA}=0.9$ of the used objective lens. Generally out-of-plane magnetization yields a higher Kerr-rotation than in-plane magnetization. During measurement skyrmion visualization results from the polar MOKE signal, imaging the out-of-plane magnetization.\\
All shown Kerr microscopy images were averaged from $16$ images. Background subtraction of $80$ times averaged images recorded for saturated magnetizations in both directions removes non-magnetic and constant-brightness artifacts from the acquired images. THE measurements are not influenced by this subtraction step and therefore it was only used in preparation for the particle detection algorithm. Additional artifacts, e.g., the Faraday effect \cite{Marko2015}, causing polarization rotation during propagation through the microscope lenses due to magnetic stray fields, are corrected in first approximation for the linear contribution ensuring a constant image brightness for both saturated magnetization states. Possible small brightness drifts resulting from the setup's unevenness are constantly corrected linearly over time. The goal of the correction is to preserve a constant brightness level throughout a hysteresis loop measurement for equal magnetic fields while imaging a saturated magnetization.\\
All these corrections ensure that intensity changes in the images are directly proportional to the surface magnetization. This configuration allows using the Kerr images, recorded while sweeping the out-of-plane applied field, to reconstruct the average magnetization from the overall brightness and extract the hysteresis loops. This surface magnetization information allows to remove all AHE contributions included in \autoref{GesHallFormel} to separate the remaining OHE and THE. At the applied field amplitudes below $0.5\,$\si{\milli\tesla}, the OHE contributions are very small, at $\sim 10^{-4}\,$\si{\nano\ohm\meter} estimated with reference \cite{Gerber2017}, and can therefore be neglected for the analysis when considering the THE peaks in \autoref{TopologicalEffectComparisonNewHall}. A significant influence of the OHE would be evident in the THE measurement data shown in \autoref{HysteresisComparisonNewHall} as a slope in the saturation regions and is not found.\\
The magnetization of the employed CoFeB films saturates at applied fields above $3\,$\si{\milli\tesla}, and all magnetic structures disappear. The resulting constant MOKE signal in magnetic saturation allows normalizing the MOKE brightness to the Hall resistivity and extracting the THE by subtracting both normalized signals. The THE resistivity is calculated from the Hall resistivity normalization. This evaluation procedure including the linear drift correction is applied analogously to extract the THE and TNE from the Nernst effect measurements, extracting $N^\mathrm{TQ}$ introduced in \autoref{TopologicalQuantityFormel}.\\
Multiple contributions from overlapping AHE effects arising from local thickness variation, defects, or interface-related defects can sum up appearing like THE peaks in the Hall effect measurement loops like e.g. in \autoref{SumAHEu10THEPlot} and are investigated in more detail in reference \cite{Kimbell2022}. Our subtraction method eliminates those contributions, strongly reducing these contributions and enabling a more reliable extraction of the THE, since the hysteresis extracted from the Kerr images also includes the AHE contributions.

\newpage
\section{Results and discussion}

\subsection{Topological Hall effect and topological quantity}
\begin{figure}[ht]
\begin{center}
\begin{subfigure}{0.495\textwidth}
  \centering
  \includegraphics[width=\textwidth]{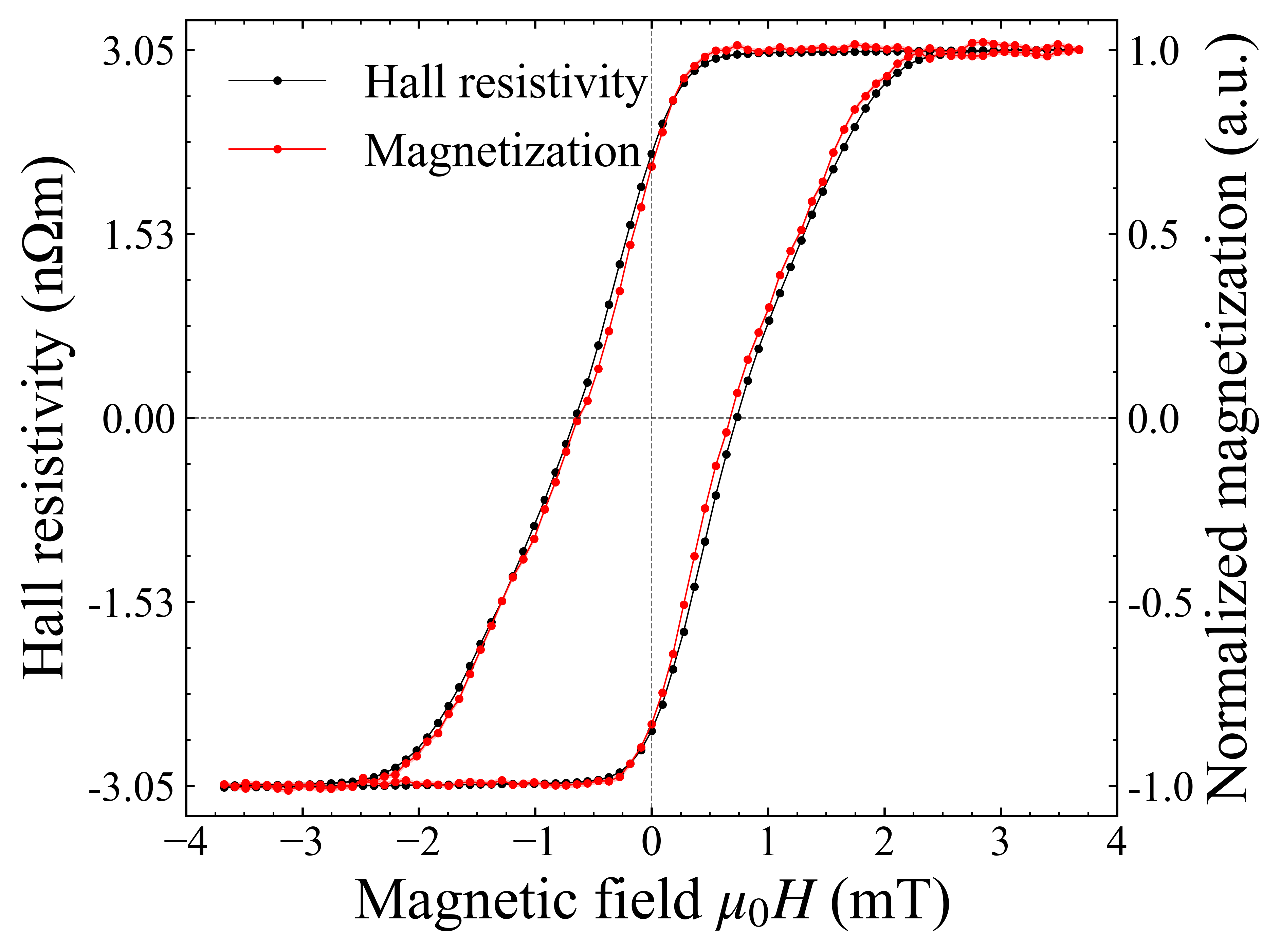}
  \caption{MOKE-Hall measurement}
  \label{HysteresisComparisonNewHall}
\end{subfigure}
\hfill
\begin{subfigure}{0.495\textwidth}
  \centering
  \includegraphics[width=\textwidth]{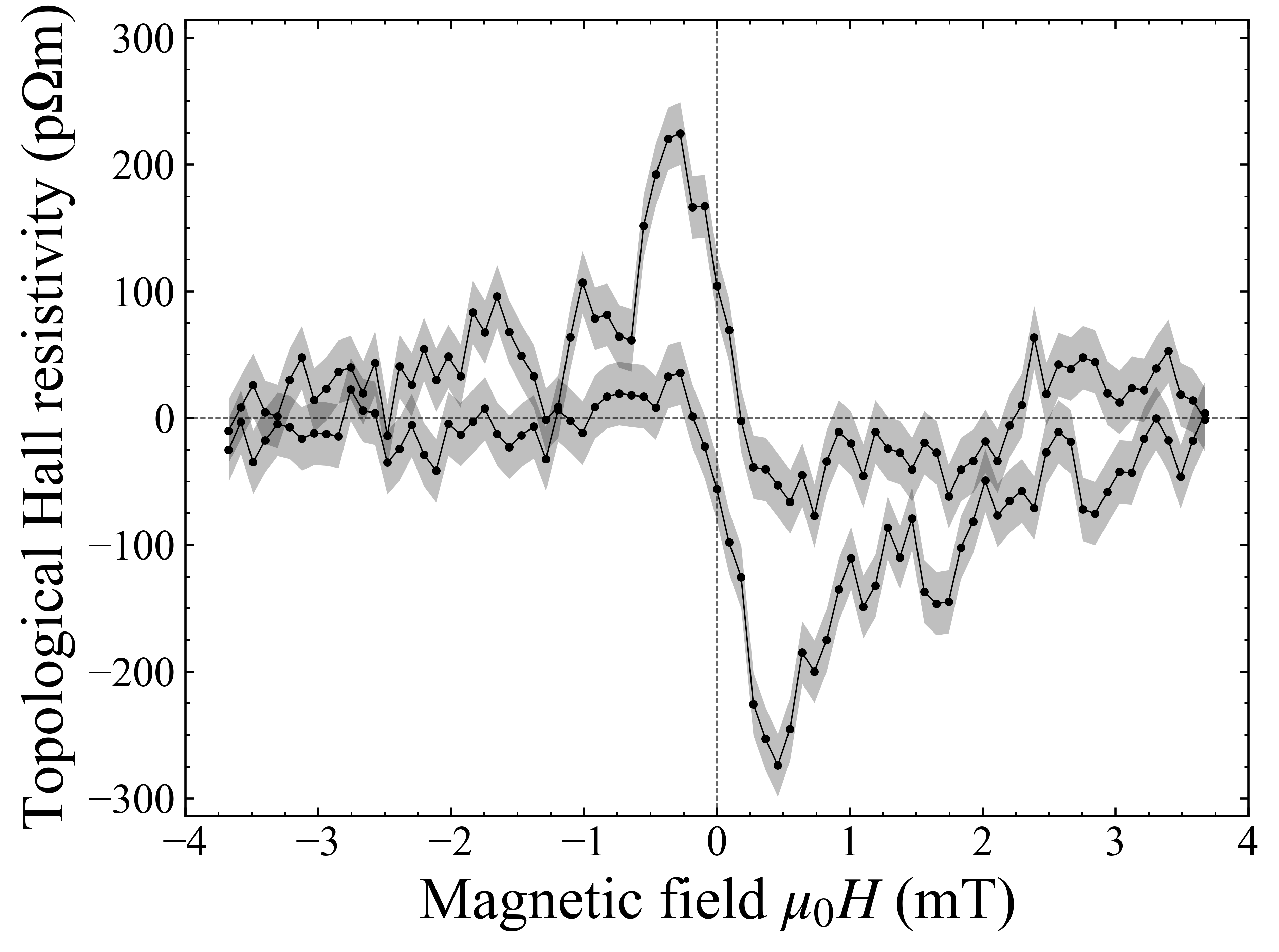}
  \caption{Topological Hall effect}
  \label{TopologicalEffectComparisonNewHall}
\end{subfigure}
\begin{subfigure}{0.495\textwidth}
  \centering
  \includegraphics[width=\textwidth]{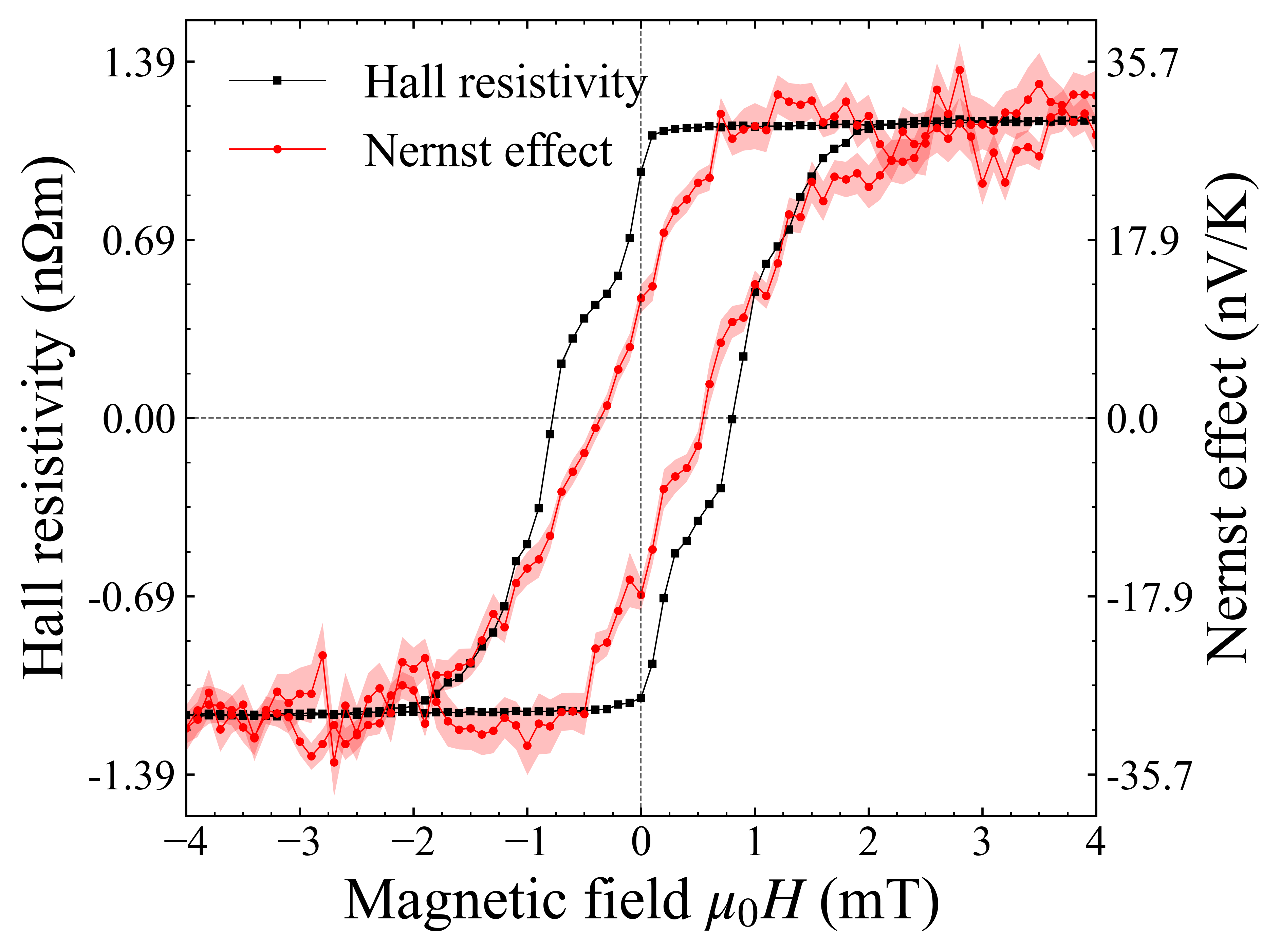}
  \caption{Nernst-Hall measurement}
  \label{HysteresisComparisonNewNernst}
\end{subfigure}
\hfill
\begin{subfigure}{0.495\textwidth}
  \centering
  \includegraphics[width=\textwidth]{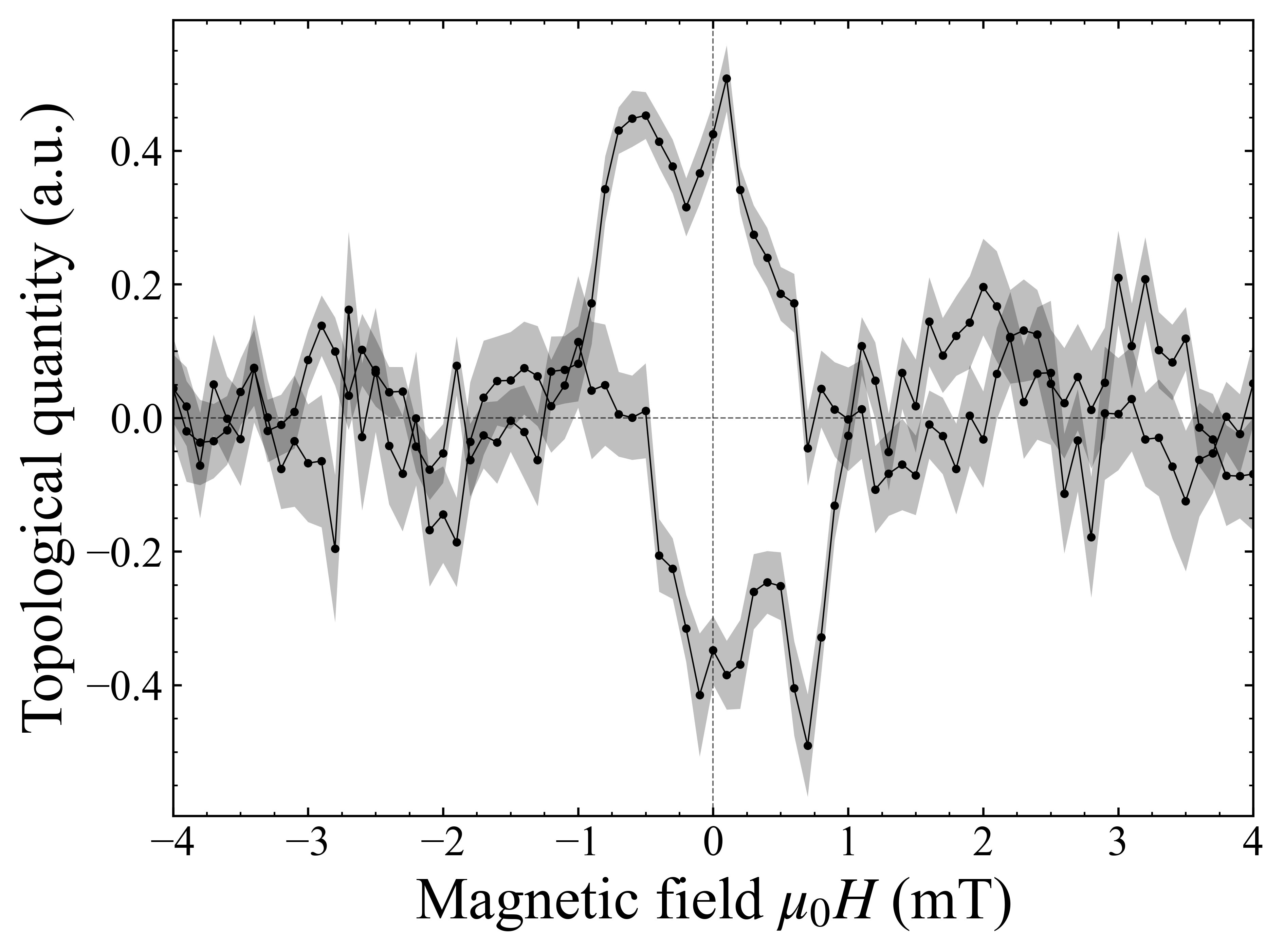}
  \caption{Topological quantity}
  \label{TopologicalEffectComparisonNewNernst}
\end{subfigure}
\caption{(a) Relative normalized hysteresis loops of the Hall resistivity (black) and magnetization (red) measured at the MOKE-Hall sample. (b) THE, is calculated from the difference of the normalized Hall resistivity and the magnetization presented in (a). (c) The Hall resistivity (black) and Nernst effect (red) measured at the Nernst-Hall sample. (d) Topological quantity $N^\mathrm{TQ}$ calculated from (c) by \autoref{TopologicalQuantityFormel}. The results in Figure (a),(b) and (c),(d) differ in the exact shape, because of slightly different positions on the CoFeB thickness wedge chosen, but reveal the quantitatively same behavior by independent experimental realization. Both experiments prove for the same related topological features.}
\label{VierPlotsHysteresenTHEMag}
\end{center}
\end{figure}

\noindent We discuss the premises for the occurrence of the skyrmion topological Hall (THE) in CoFeB thin films in the transition region from in-plane to out-of-plane anisotropy is discussed, based on the experimental. The Hall resistivity is calculated according to \autoref{HallResistivityFormel}. To extract the THE, the Kerr-derived magnetization is normalized to the Hall resistivity in saturation and subtracted from the measured Hall-resistivity loop. Since the OHE is negligible within the applied field range, the remaining difference is attributed to the THE.\\
\autoref{HysteresisComparisonNewHall} depicts the resistivity hysteresis loop obtained from the Hall effect measurement (black, left scale) and the normalized magnetization hysteresis loop extracted from the simultaneously recorded Kerr images (red, right scale). At first glance, a difference between the two loops is only visible when the magnetization departs from saturation and after the external field reverses its sign, within a field range up to $0.7\,$\si{\milli\tesla}. The Kerr images in \autoref{KerrBildSkyrmionenGesamt} confirm the presence of skyrmions within this narrow field range of around $0.7\,\si{\milli\tesla}$. Possible artifacts induced through relaxation effects and their exclusion are discussed below.\\
\autoref{TopologicalEffectComparisonNewHall} highlights the THE obtained from the difference between the two loops from \autoref{HysteresisComparisonNewHall}. The two extracted peaks reverse with the applied field, i.e., they are dependent on the initial out-of-plane saturation magnetization direction and show only a slight asymmetry within the measuring noise. The measurements reveal an averaged topological Hall resistivity $\rho^\mathrm{THE}=249(18)\,$\si{\pico\ohm\meter} for a single-layer Ta/CoFeB/MgO film, which is one order of magnitude lower than measured for Ta/CoFeB/MgO multilayers in reference \cite{A_Lone2024}.\\
We verify the results obtained using this method and the existence of topologically protected magnetic structures employing an additional experiment. The above-introduced samples (\autoref{SampleStructureTNEHall} and \ref{SampleStructureTNENernst}), patterned for thermal transport experiments, contain the same Ta/CoFeB/MgO layers with a transition region for anisotropy landscapes exhibiting skyrmions. Temperatures achieved through heating during the measurement do not exceed $12\,\si{K}$ above room temperature.\\
\autoref{HysteresisComparisonNewNernst} depicts the Hall effect (black, left scale) and the Nernst effect (red, right scale) hysteresis loops measured at the same magnetic conditions in an area exhibiting skyrmions. Both hysteresis loops exhibit similar slope differences, as in the data from \autoref{HysteresisComparisonNewHall}, for applied field ranges at which skyrmions emerge. The resulting $N^\mathrm{TQ}$ is calculated from the difference between the two loops by \autoref{TopologicalQuantityFormel} and is plotted in \autoref{TopologicalEffectComparisonNewNernst}. This data clearly displays the characteristic peaks, as visible in the data plotted in \autoref{TopologicalEffectComparisonNewHall}, which vary in width and form.\\
The material at the observed sample positions varies by only a few picometers within the CoFeB gradient along the wedge influencing the anisotropy, which sets the conditions for skyrmion formation. This circumstance complicates finding identical conditions for skyrmion generation in two different samples. In reference \cite{Denker2023} those conditions are examined closely for thin CoFeB films, showing how a small controlled thickness variation along the growth gradient shifts the anisotropy landscape and the required applied field magnitudes to create skyrmions. Hence, the results in \autoref{HysteresisComparisonNewHall},\subref{TopologicalEffectComparisonNewHall} and \ref{HysteresisComparisonNewNernst},\subref{TopologicalEffectComparisonNewNernst} differ quantitatively. The two selected examples in \autoref{VierPlotsHysteresenTHEMag} feature slightly different skyrmion formation field ranges in their phase diagram, controlled by the Ta/CoFeB/MgO thickness wedge structures, highly sensitive to the CoFeB thickness. Both effects, the topological Hall (THE) and Nernst effect (TNE), result from spin structures, such as skyrmions, and are summed up in the topological quantity $N^\mathrm{TQ}$. This makes the measured effect in $N^\mathrm{TQ}$ larger compared to the pure THE measured by MOKE and Hall resistivity, but does not change its overall structure in the measured data. A direct quantitative comparison of the THE in \autoref{TopologicalEffectComparisonNewHall} and the $N^\mathrm{TQ}$ of \autoref{TopologicalEffectComparisonNewNernst} is not possible in those separate measurements, because the TNE units (\si{\volt}/\si{\kelvin}) in $N^\mathrm{TQ}$ are not comparable with THE units (\si{\ohm\meter}). In contrast to the MOKE-Hall measurement, it is not possible to separate the THE from the TNE in the $N^\mathrm{TQ}$ measurement, by just measuring the Nernst and Hall effects. The Hall resistivity in saturation in \autoref{HysteresisComparisonNewHall} and \subref{HysteresisComparisonNewNernst} should be the same for both measurements and samples, but they differ by a factor of approximately $2.6$. The slightly different CoFeB thicknesses and therefore the magnetic anisotropy of both samples cause this discrepancy. They are especially evident in the hysteresis loops.\\
The $N^\mathrm{TQ}$ measurement confirms the existence of the THE and TNE in Ta/CoFeB/MgO and therefore validates the method of the THE measurement by comparing MOKE and Hall resistivity signals.

\newpage
\subsection{Observation of the magnetic structures during THE measurement}

\begin{figure}[ht!]
\begin{center}
\begin{subfigure}{0.325\textwidth}
  \centering
  \includegraphics[width=\textwidth]{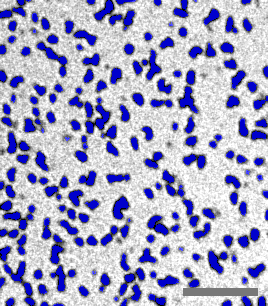}
  \caption{Rising edge\\ at $\mu_0H=0.09(5)\,$\si{\milli\tesla}}
  \label{KerrBildSkyrmionenA}
\end{subfigure}
\hfill
\begin{subfigure}{0.325\textwidth}
  \centering
  \includegraphics[width=\textwidth]{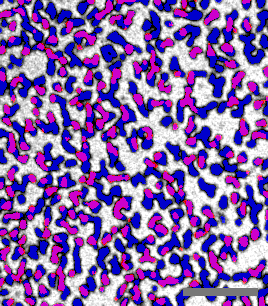}
  \caption{Peak\\ at $\mu_0H=-0.28(5)\,$\si{\milli\tesla}}
  \label{KerrBildSkyrmionenB}
\end{subfigure}
\hfill
\begin{subfigure}{0.325\textwidth}
  \centering
  \includegraphics[width=\textwidth]{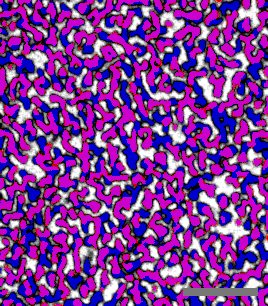}
  \caption{Falling edge\\ at $\mu_0H=-0.64(5)\,$\si{\milli\tesla}}
  \label{KerrBildSkyrmionenC}
\end{subfigure}
\begin{subfigure}{0.7\textwidth}
  \centering
  \includegraphics[width=\textwidth]{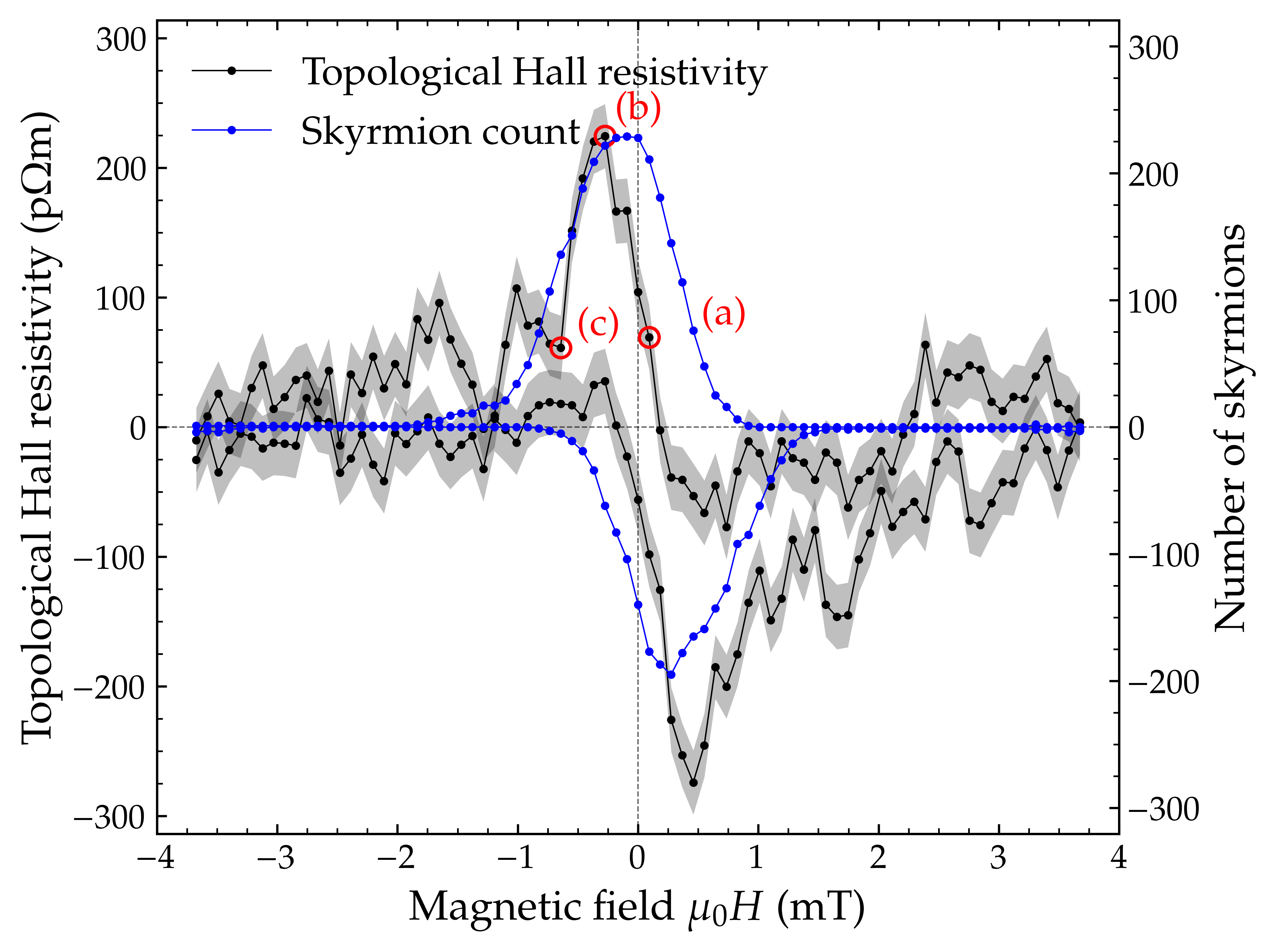}
  \caption{Comparison of THE and number of skyrmions}
  \label{THEMagPlotVergleich}
\end{subfigure}
\caption{Kerr images of magnetic structures with colorized skyrmion positions (a-c) at the red marked points of THE peak of (d) when starting at positive magnetization saturation. The scale bars correspond to $5\,$\si{\micro\meter}. (a) Is the the rising edge on the right side and (c) is falling edge on the left side with the peak (b) in the middle. All skyrmions detected in the pictures are marked in blue and for comparison (b) shows the skyrmion positions of (a) in red, while (c) shows the positions of (b) in red. The purple regions mark the overlap of skyrmions from the previous position. The out-of-plane magnetization in brighter areas points upwards and in darker/colorized areas downwards. (d) shows a comparison between the THE and number of skyrmions. Its sign is changed for one half of the measurement for better comparability with the THE signal.}
\label{KerrBildSkyrmionenGesamt}
\end{center}
\end{figure}

\noindent Figures \ref{KerrBildSkyrmionenA}-\subref{KerrBildSkyrmionenC} show exemplary Kerr microscopy images, recorded during THE measurement, chosen for further investigation to identify the skyrmion formation. Its structures and particle counts are investigated to find the origin of the observed THE peak signal plotted in \autoref{THEMagPlotVergleich}.\\
The number of skyrmions at \autoref{THEMagPlotVergleich} are calculated from Kerr microscopy images though particle detection using a threshold filter algorithm in Python \cite{PSF} with OpenCV \cite{OpenCV} and SciPy \cite{2020SciPyNMeth}. Its image preparation for particle size calculation includes applying a difference of Gaussian (DoG) filter \cite{Lowe2004} for improving the particle separation and detection. It works by subtracting a Gaussian blur image with a sigma of $4\,$pixel multiplied with the factor $0.7$ from the original image. Subsequently, a $0.6\,$pixel Gaussian blur filter is applied for noise reduction. The threshold values are set manually for images taken at both THE peaks and applied to both hysteresis measurement loop directions separately. Within the particle detection algorithm, only particles larger than $4\,\mathrm{pixels}$ are considered in order to exclude image noise that may be misidentified as particles.\\
The particle number peaks, detected with Kerr microscopy, align well with the THE peak in \autoref{THEMagPlotVergleich} and almost overlap with the THE following the leading particle number peaks. These particles are skyrmions as expected for this material \cite{Denker2023}, while deformations of the skyrmion forms are caused by the specific energy landscape \cite{Gruber2022}. The deformation shows how the object probe their magnetic surrounding. Its peak heights do not align, but still allow for the estimation of a THE per skyrmion of $\approx 0.1\,\si{\pico\ohm\meter}/\mathrm{skyrmion}$ when considering the measured number of skyrmion at the THE peaks and extrapolate it from the field of view to the relevant Hall cross area.\\
In \autoref{KerrBildSkyrmionenGesamt} the THE signal is increasing at the rising edge, maximal at the peak and decreasing at the falling edge. At the rising edge, shown in \autoref{KerrBildSkyrmionenA}, the skyrmion size and density is small but increasing to the peak in \autoref{KerrBildSkyrmionenB} until the structures grow into larger magnetic structures at the falling edge in \autoref{KerrBildSkyrmionenC} and the THE and number of skyrmions decrease again. It shows that the visible skyrmions are likely causing the THE signal until they get to large and few, while larger structures, resulting from skyrmion merging, contribute decreasingly to the THE signal. This indicates that the extracted THE does not necessarily scale only with the number of detected skyrmions once the system leaves the regime of isolated skyrmions. In the dense or partially merged multidomain state, the Hall response also depend on the domain morphology, domain wall length, stripe-domain end caps, and topological defects, which are known to influence transitions between labyrinthine domains and skyrmions in Ta/CoFeB/MgO \cite{Wang2019}. Moreover, multidomain ferromagnets can produce THE-like Hall anomalies through AHE and domain-wall-scattering effects alone, even without a finite topological charge \cite{Sabri2025}. Therefore, the comparison between Kerr microscopy and the extracted THE should be understood as evidence for a correlation between the observed magnetic textures and the transport signal, rather than as a strict one-to-one proportionality between THE amplitude and number of skyrmions.

\subsection{Magnetic relaxation}
\label{MagneticRelaxationAppendix}

\begin{figure}[ht]
\begin{center}
\begin{subfigure}{0.3\textwidth}
  \centering
  \includegraphics[width=\textwidth]{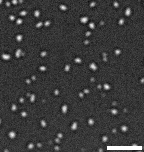}
  \caption{}
  \label{MOKEKriechenBildStart}
\end{subfigure}
\begin{subfigure}{0.3\textwidth}
  \centering
  \includegraphics[width=\textwidth]{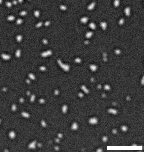}
  \caption{}
  \label{MOKEKriechenBildEnde}
\end{subfigure}
\begin{subfigure}{0.3\textwidth}
  \centering
  \includegraphics[width=\textwidth]{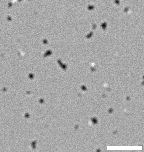}
  \caption{Difference of (a) and (b)}
  \label{MOKEKriechenBildDiff}
\end{subfigure}
\caption{Magnetic creep of skyrmions investigated with Kerr imaging in (a) and (b). The skyrmions were nucleated originating from negative saturation with a series of in-plane and out-of-plane magnetic fields followed by a constant stabilization field of $-0.92(5)\,$\si{\milli\tesla} at a sample similar to the MOKE-Hall sample. (a) was measured $4\,$\si{\second} after skyrmion nucleation. The waiting time between (a) and (b) is around $25\,$\si{\second}. (c) shows the difference image of (a) and (b). The scale bars correspond to $5\,$\si{\micro\meter}.}
\label{MOKEKriechen}
\end{center}
\end{figure}

\begin{figure}[ht]
\begin{center}
\includegraphics[width=0.65\textwidth]{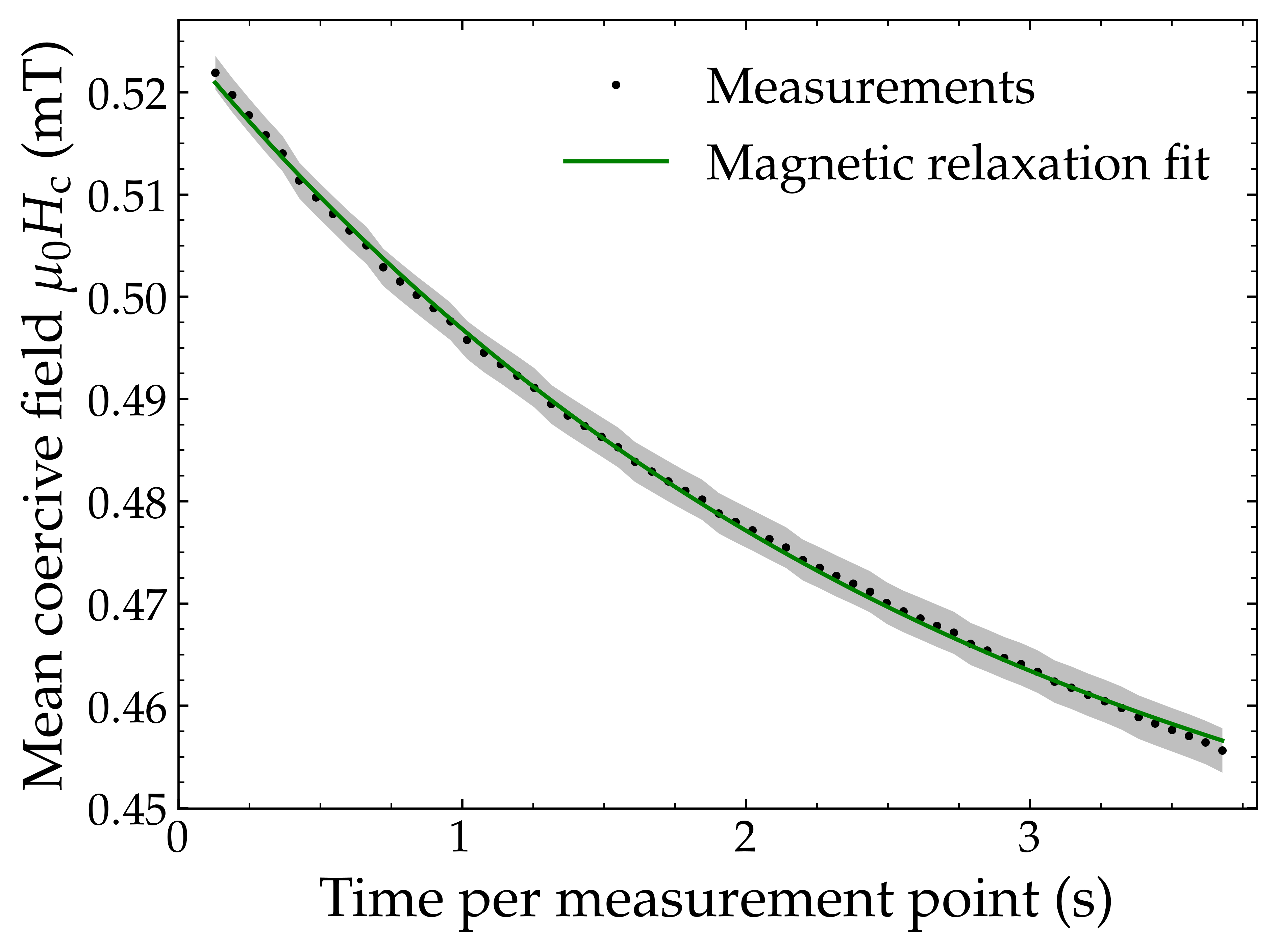}
\caption{Magnetic relaxation at the coercive field for a typical CoFeB skyrmion layer. The measurement procedure involves applying the field values and subsequently recording the Hall voltage over the time range $0.1\,\si{\second} < t < 3.7\,\si{\second}$, while completing a hysteresis loop. At each measurement point of the hysteresis loop, $61$ subsequent measurements were acquired. The measurement is repeated and averaged $32$ times.}
\label{PlotMagneticRelaxation}
\end{center}
\end{figure}

\noindent Deformation effects of topological objects, together with magnetic creep caused by domain nucleation, propagation, and anisotropy \cite{Boukari2001}, both induce drifts in the measured hysteresis data. Comparing hysteresis loops obtained in different measurements, such drifts can produce signatures similar to the peaks we extract to determine the topological Hall effect and the topological Nernst effect. Especially in regions with a large slope, which are essential in our analysis, such artifacts must be excluded. \autoref{MOKEKriechen} shows two unprocessed images, recorded $4\,$\si{\second} after skyrmion nucleation (\subref{MOKEKriechenBildStart}) and $25\,$\si{\second} after nucleation (\subref{MOKEKriechenBildEnde}). \autoref{MOKEKriechenBildDiff} shows the difference between the two previous images. The remaining dark and bright spots indicate a deformation of the skyrmions over the recorded time span. We ascribe this slow deformation effect to thermal fluctuations, generating a negligible impact on the shape of the hysteresis loop. We account for the magnetic creep in our analysis by determining the time scales of the magnetic relaxation. In \autoref{PlotMagneticRelaxation}, we examine the coercive field as an indicator for magnetic relaxation because within this region, the magnetization changes with the applied field are large. The values are extracted from a hysteresis loop measured by setting the applied field amplitude and recording the values after a waiting time $0.1\,\mathrm{s} < t < 3.7\,\mathrm{s}$. This magnetic relaxation follows an exponential Arrhenius-type relaxation and can be extracted from coercive fields by:
\begin{equation}
\mu_0 H_{c}=A + B\mathrm{e}^{-t/t_0}.
\label{DecayFitFormel}
\end{equation}
Here, $t$ is the waiting time after field application and prior to measurements, $A$ the coercive field approached by the the magnetic relaxation with the relaxation amplitude $B$ and $t_0$ the relaxation time with the half life $t_{1/2}=t_0\cdot \ln(2)$. From fitting result the parameters $A=0.4322(19)\,$\si{\milli\tesla}, $B=0.0930(17)\,$\si{\milli\tesla}, $t_0=2.75(11)\,$\si{\second} and $t_{1/2}=1.90(8)\,$\si{\second}. After a waiting time of $t=0.5\,$\si{\second}, relaxation processes are mainly decayed, and changes in magnetization are decreasing to a negligible level. Therefore, it is sufficient to begin recording the Hall voltage for $1.3\,$\si{\second} after the initial waiting time $t=0.5\,$\si{\second}, and record the Kerr images subsequently within a time span of $1.9\,$\si{\second}. We use this time constant between applying the field and recording the data points throughout the measurements. The data confirm that, over the time span used across the presented experiments, the relaxation up to the coercive field varies by less than one applied-field point of $0.1\,$\si{\milli\tesla}, which is within the error range. Additionally, both peaks, the skyrmion count and the topological Hall effect, match at approximately the same applied field. The skyrmion count is extracted from the Kerr images only and is therefore independent of the Hall-voltage measurement. A significant magnetization relaxation effect would produce a mismatch of several applied-field points between the two peaks. If magnetic relaxation effects remain, they may weaken the topological Hall effect in the measurements, but they do not generate detectable artifacts. For the independent $N^\mathrm{TQ}$ measurements the relaxation is irrelevant, because its Nernst and Hall effect measurements were conducted with the same relaxation states.\\
Similar to \autoref{DecayFitFormel} an exponential dependency of the coercive field on the time $t$ was modeled in \cite{Vopsaroiu2010,Cui2022} for ferroelectrics. The simple model in reference \cite{Vopsaroiu2010} assumed domain nucleation as the relevant reversal mechanism for the hysteresis curve. Since the underlying physical processes are observed in many other systems, we postulate that the main mechanism is widely applicable to many hysteretic systems and can provide a lowest order approximation for the coercive field in our experiments, as shown in our \autoref{DecayFitFormel} and its agreement with our experimental data shown in \autoref{PlotMagneticRelaxation}. The model of \cite{Vopsaroiu2010} predicts that the coefficient $B$ is proportional to the temperature $T$, i.e. if the measurement time is fixed the coercive field should decrease linearly with temperature. Future experiments varying temperature could test this prediction. Further investigations of the magnetic relaxation in the context of skyrmion host materials may reveal insights to skyrmion stability.\\

\section{Summary}
We investigate the topological Hall effect in electrical transport measurements combined with Kerr microscopy and the topological Nernst effect in thermal transport measurements at room temperature to determine skyrmion nucleation in the single-layer nanometer thick magnetic material Ta/CoFeB/MgO stacking. Both methods independently enable the separation of topological effects from the total Hall and Nernst effects by showing peaks during skyrmion nucleation in the difference signal.\\
Kerr microscopy allows us to isolate the topological Hall effect from the total Hall effect, while simultaneously investigating the formation of skyrmions optically with a high spatial resolution. We identify the deformation and magnetic relaxation of these topological objects. The measured topological Hall effect peaks at a mean peak height of $248(18)\,$\si{\pico\ohm\meter} and its trend agrees with the detected number of skyrmions, indicating their topology as the origin of this effect, even though a significant deformation and magnetic creep is found.\\
The thermal transport measurement supports this finding without relying on Kerr microscopy. Therefore we can verify this by two independent methods. In summary, the measurements in this work prove the existence of topological magnetic skyrmions in the given material and demonstrate the disentanglement of their hidden topological effects.

\section{Funding}
Funding by the German Research Foundation (DFG) through the grant MC 9/22-1 (Project No. 438892000) for F. Gossing and VO 3017/1-1 (Project No. 571878852) for Dr. M. Vogel is gratefully acknowledged.

\bibliography{THEinTaCoFeBMgOLiteratur}
\end{document}